\pdfoutput=1
\documentclass[10pt,aps,prr,amsmath,amssymb,twocolumn,raggedbottom,
letterpaper,final,superscriptaddress,citeautoscript,floatfix]{revtex4-1}
\usepackage[usenames,dvipsnames]{color}
\usepackage{graphicx,microtype}
\usepackage[bookmarks=false,colorlinks]{hyperref}
\hypersetup{
    linkcolor=magenta,        
    citecolor=MidnightBlue,     
    filecolor=Plum,      	
    urlcolor=MidnightBlue,           
}

\newcommand{\CRO}{Ca$_3$Ru$_2$O$_7$}
\newcommand{\dpg}[1]{\textcolor{black}{#1}}



\begin{document}
\title{Cooperative Interactions Govern the Fermiology of the Polar Metal \CRO}

\author{Danilo Puggioni}
\email{danilo.puggioni@northwestern.edu}
\affiliation{Department of Materials Science and Engineering, Northwestern University, Evanston IL, 60208, USA}

\author{M.\ Horio}
\affiliation{Physik-Institut, Universit\"at Z\"urich, Winterthurerstrasse 190, CH-8057 Z\"urich, Switzerland}

\author{J.\ Chang}
\affiliation{Physik-Institut, Universit\"at Z\"urich, Winterthurerstrasse 190, CH-8057 Z\"urich, Switzerland}

\author{James M.\ Rondinelli}
\email{jrondinelli@northwestern.edu}
\affiliation{Department of Materials Science and Engineering, Northwestern University, Evanston IL, 60208, USA}

\begin{abstract}
The antiferromagnetic Ruddlesden-Popper ruthenate \CRO\ is a model polar metal, combining inversion symmetry breaking with 
metallic conductivity; however, its low temperature ($T < 48$\,K) crystal structure and Fermi surface topology 
remain ambiguous despite numerous measurements and theoretical studies. 
Here we perform both first principles calculations with static correlations 
and angle resolved photoelectron spectroscopy experiments to construct a 
complete model of \CRO, reconciling inconsistencies among interpretations of 
electrical transport, thermopower measurements, and momentum- and energy-resolved 
band dispersions.
The solution relies on treating the interplay among Coulomb repulsion, 
magnetic ordering, spin-orbit interactions, and the RuO$_6$ octahedral degrees-of-freedom on equal footing.
For temperatures $30<T < 48$\,K, we propose weak electron-electron interactions produce a symmetry-preserving 
metal-semimetal transition with Weyl nodes in proximity to the Fermi level, whereas 
a new orthorhombic $Pn2_1a$ structure emerges for $T<30$\,K,
exhibiting charge and spin density waves from enhanced Coulombic interactions. 
\end{abstract}

\date{\today}

\maketitle

\section{Introduction}
\begin{figure}
\centering
\includegraphics[width=0.95\columnwidth]{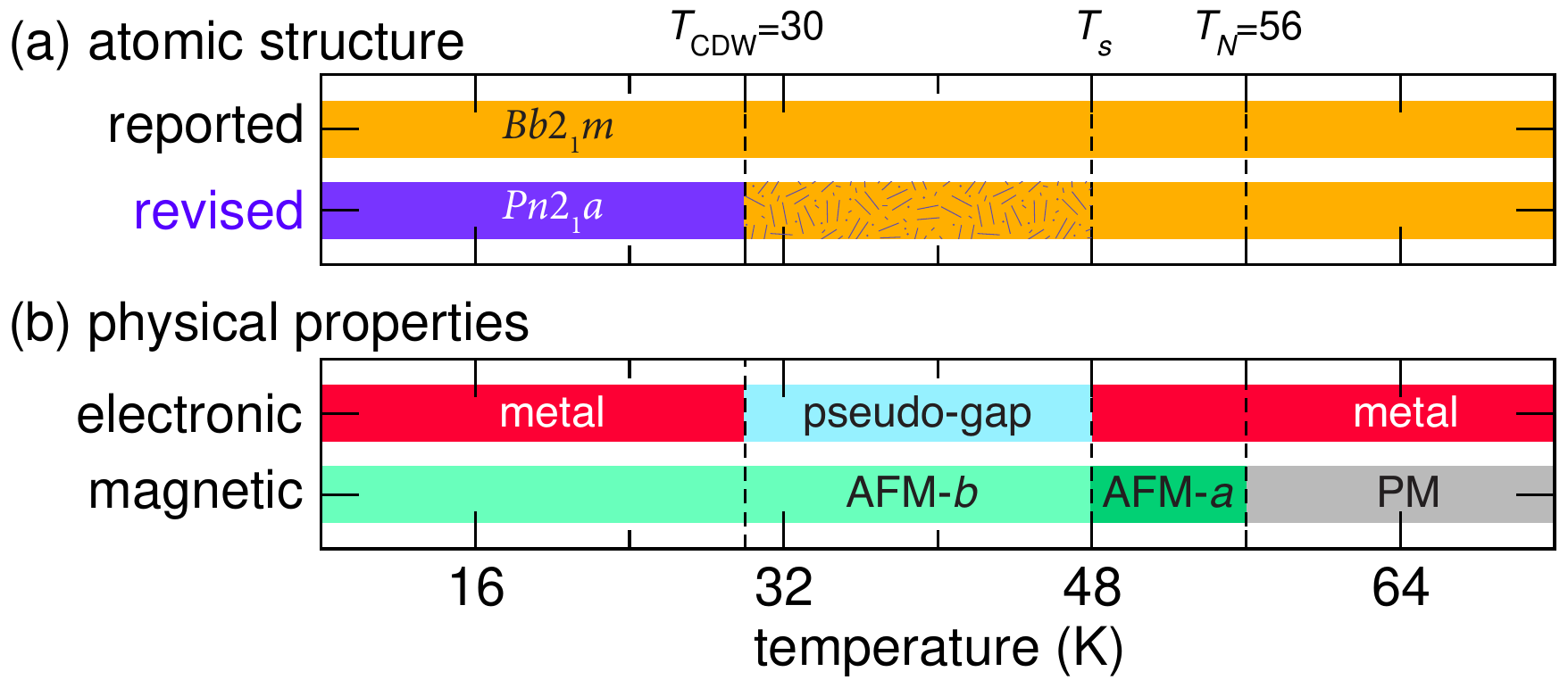}\vspace{-10pt}
  \caption{Low-temperature phase diagram 
  of Ca$_3$Ru$_2$O$_7$, which assigns isosymmetric (symmetry-preserving) transitions to the magnetic ($T_N$ and $T_s$) and electronic $T_\mathrm{CDW}$ transitions according to Ref.\ \onlinecite{yoshida2005}. The revised diagram replaces the $Bb2_1m$ symmetry with $Pn2_1a$ for $T<T_\mathrm{CDW}$; the shaded region indicates that the $Pn2_1a$ may persist to $T_s$.}
 \label{CRO_phase_diag}
\end{figure}

\begin{figure*}
\centering
\includegraphics[width=0.95\textwidth]{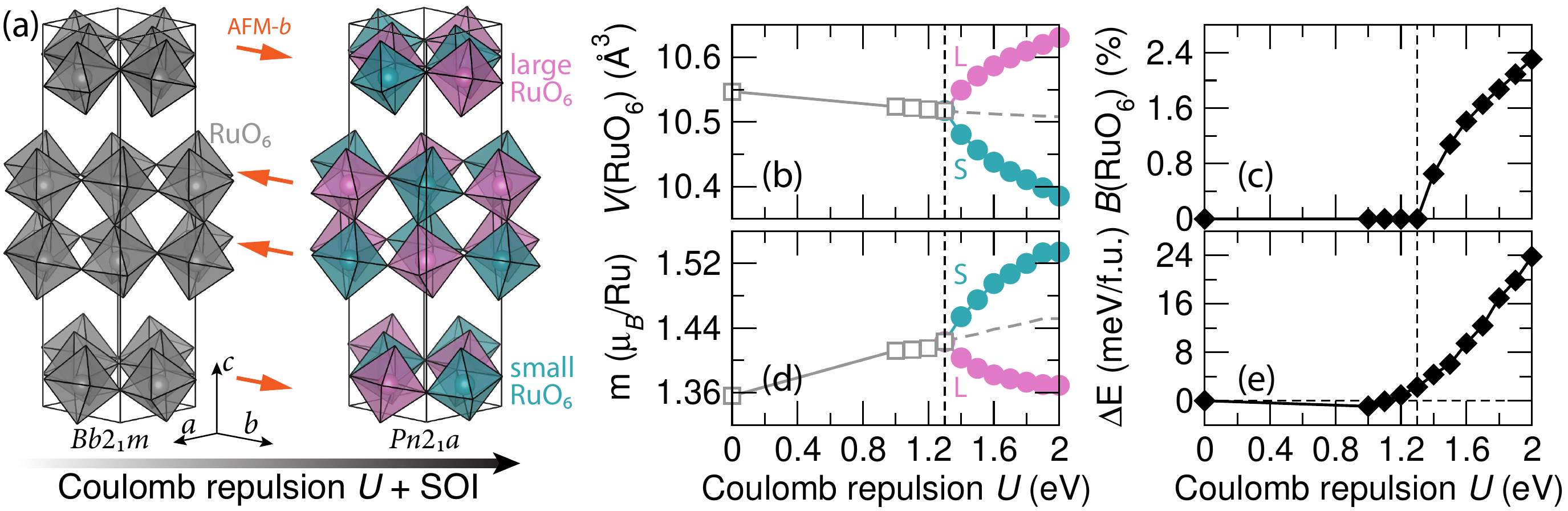}
\caption{(a) Crystal structure of  \CRO\ as a function of the Coulomb repulsion ($U$) with calcium and oxygen atoms omitted for clarity. 
For $U>1.3$\,eV, ordering of dilated (large, L) and contracted (small, S) RuO$_6$ octahedra
occur in a rock-salt-like  pattern, reducing the symmetry from $Bb2_1m \rightarrow Pn2_1a$.
Change in 
(b) RuO$_6$ octahedra volume, $V$, 
(c) percent difference in volumes of two adjacent octahedra, $B(\mathrm{RuO}_6)$, and
(d) local Ru spin moment with $U+\mathrm{SOI}$. 
(e) Coulomb-dependent energy difference ($\Delta E$) between the $Bb2_1m$ and $Pn2_1a$ phases 
without SOI. 
}
\label{fig1}
\end{figure*}

Ruddlesden-Popper (RP) \CRO\ is a unique polar metal exhibiting a rich phase diagram (\autoref{CRO_phase_diag}), 
colossal magnetoresistance \cite{Cao_Agterberg:2003,Lin_Cao:2005}, 
highly anisotropic electrical resistivity \cite{Cao_2004}, and  polar domain 
structures \cite{shiming:2018,*Stone:2019}.
It displays multiple magnetic and electronic transitions 
without changes in crystal symmetry, making it also a member of 
a rare condensed-matter family exhibiting isosymmetric transitions \cite{Christy1995}. 
\CRO\ exhibits the polar space group $Bb2_1m$ 
from room temperature (RT) to 8\,K 
\cite{yoshida2005} owing to in-phase and out-of-phase RuO$_6$ octahedral rotations along the [001] and [110] directions, respectively 
\cite{shiming:2018}.  
Below $T_N=56$\,K, it  exhibits antiferromagnetic order (AFM-$a$) characterized by 
ferromagnetic perovskite bilayers with a [100] easy axis, \emph{i.e.}, 
the short axis of the crystal.
The bi-layers then couple antiferromagnetically along [001].
At $T_s=48$\,K, it undergoes a first-order isosymmetric 
transition marked by 
a discontinuous change in lattice parameters but 
no change in occupied Wyckoff positions or space group. 
Interestingly below $T_s$, the lattice constants are nearly temperature independent   \cite{yoshida2005}, despite an electrical resistivity transition from a three-to-two-dimensional state. 
Coincident with $T_s$, a second magnetic transition occurs, AFM-$a\rightarrow$AFM-$b$; the easy axis switches from the [100] to [010] direction (\autoref{fig1}a).
Understanding how multiple orders harmonize and evolve with broken 
inversion and time reversal symmetries as in \CRO\ is critical 
for control of topological quasi-particle excitations  \cite{Basov2017}. 

Angle resolved photoelectron spectroscopy (ARPES)  and optical-pump/optical-probe experiments propose an electronic instability underlies the transport anisotropy 
at $T_s$ by removal of well-defined electron- and hole-like pockets on the Fermi surface (FS) upon cooling 
\cite{Baumberger:2006,spinreorientation,Yuan:2019,ARPES2019}.
The electron-like band about the $\Gamma$ point is removed and asymmetric FS changes 
occur at the zone boundary \cite{ARPES2019}: 
A temperature-dependent electron pocket at \dpg{$k=$~M$(\pi/a,0)$} 
shrinks upon cooling through $T=30$\,K while 
a temperature independent hole pocket is found at \dpg{M$^\prime(0,\pi/b)$}.
These data suggest either a FS reconstruction \cite{ARPES2019} 
or the appearance of a charge density wave at $T_s$  \cite{Leejs:2007}, respectively; 
however, there remains no direct evidence for its presence in \CRO. 
Below 30~K, quasi-2D 
electron and hole pockets (areas $\approx$0.3\,\% of the Brillouin zone) survive and form the low-temperature FS as determined from 
quantum oscillations  \cite{Baumberger:2006}.

In contrast,  thermoelectric measurements show strong asymmetric FS changes 
with temperature based on assignments of the principal 
carrier-type with the sign of the thermopower coefficient along the $a$ and $b$ crystal axes \cite{Xing:2018}.
This transport-deduced evolution is inconsistent with the aforementioned optics.  
Above $T_s$, only hole-like bands are 
reported, whereas only electronic-like bands contribute for 
intermediate temperatures, $30<T<48$\,K, in disagreement with the ARPES resolved  
electron pocket dependencies about \dpg{$k=$~M$(\pi/a,0)$}. 
Below $T=30$\,K, however, negative and positive 
thermopower responses were measured along the crystallographic $a$ and $b$ axes, respectively, 
indicating predominately electron- and hole-like bands along
these directions on the FS---features more consistent with the ARPES. 
Interestingly, all  previous band-structure calculations based on 
density functional theory (DFT) \cite{LiuGuo-Qiang2011,Zheting:2018} are also 
inconsistent with the ARPES \cite{Baumberger:2006,ARPES2019,spinreorientation} and 
thermoelectric transport measurements \cite{Xing:2018}.
This is unusual given that DFT methods typically provide accurate descriptions of 
weakly correlated metals \cite{Guo/Rondinelli:2011, Grebinskij:2013, Etz:2012}.
This discrepancy may be due, in part, to underlying structural model assumptions
deduced on the reported nearly temperature-independent 
changes in the atomic structure below $T_s$ and difficulties in resolving weak 
reflections in layered perovskites \cite{Akamatsu:2014,zhu:2017}.

Which interactions cause the change in the electronic structure at $T_s=48$\,K and 
$T=30$\,K and how can one reconcile the aforementioned discrepancies remains unclarified.
Here we provide a model for the low temperature phases of \CRO\  
based on a single-particle DFT plus Hubbard $U$ description, which is fully 
consistent with all available data.
For all values of the Coulomb interaction $U$, we  find a 
van Hove singularity (vHs) within $\lesssim$20\,meV of the 
Fermi level owing to static Coulomb interactions that enhance broken 
band degeneracies near \dpg{$k=$~X$(\pi/a,\pi/b)$} split by the spin-orbit interaction (SOI).
The filling of these states occurs proximate to Weyl points 
in momentum space that arise from broken inversion and time-reversal symmetries.
Thus, the FS  comprises both electron- and hole-like pockets; however, 
the vHs prohibits using thermopower data alone to deduce the evolution of these bands. 
We use this electronic structure to propose that a Lifshitz-like electronic 
transition first occurs at $T_s=48$\,K  for $U<1.3$\,eV
and a gapped-out FS persists for $30<T<48$\,K within the known $Bb2_1m$ structure. 
Next upon increasing the value of $U$ (cooling below 30\,K), 
we stabilize a unique low-temperature phase ($Pn2_1a$) with charge/spin density waves that 
support conclusions of earlier optical spectroscopy experiments \cite{Leejs:2007}.
Finally, we compare our $T\leq30$\,K band structure calculations for  
$Pn2_1a$ \CRO\ with direct ARPES data to confirm both the existence of  linear band dispersions 
below the Fermi level (15~meV) at \dpg{M$(\pi/a,0)$} and that enhanced interactions ($U=1.6$~eV) with both SOI and RuO$_6$ distortions 
are required to describe the fermiology. 
Thus, we propose the ground state crystal structure is  $Pn2_1a$ and that this phase may 
persist above $T=30$\,K and below $T_s=48$\,K.

\section{Model and Methods}
We formulate a model that explicitly accounts for the interplay among  Coulomb repulsion, magnetic ordering, 
SOI, and the structural degrees-of-freedom---a combination of features 
that has not been previously pursued. 
All  previous DFT studies used the reported experimental structure 
\cite{yoshida2005}, neglecting  
atomic relaxations \cite{LiuGuo-Qiang2011, Zheting:2018,spinreorientation}.
We include static correlations through a local Coulomb interaction $U$, which is 
screened through a temperature dependent electronic bandwidth ($W$) as $U/W(T)$, 
and further stabilizes states with orbital degeneracies broken by the SOI.
We justify this approach as follows: First at $T_s$, the $c$-axis compression increases 
the RuO$_6$ octahedral rotational angles, which are known to narrow the bandwidth 
(reduce $W$) of the low-energy $t_{2g}$ orbitals  in oxides comprising 
octahedral $\mathrm{Ru}^{4+}$. These angles exhibit a further deviation from $180^\circ$ 
upon cooling \cite{yoshida2005}.
Second, finite temperature will effectively reduce the local interaction allowing 
double occupancy of the $d$ manifold, increasing toward a non-interacting 
value at high temperature \cite{PhysRevB.95.235109,*PhysRevB.98.205114}.
Thus, the effective $U$ at room-temperature will be smaller than that at 0\,K.

We perform first-principles DFT$+U$ calculations 
\footnote{%
We use the Perdew-Burke-Ernzerhof exchange-correlation functional revised for solids (PBEsol) 
\cite{PBEsol:2008} as implemented in the Vienna \textit{Ab initio} Simulation Package (VASP) 
\cite{Kresse/Furthmuller:1996b} 
with the projector augmented wave (PAW) approach \cite{Blochl/Jepsen/Andersen:1994} 
to treat the core and valence electrons using 
the following  electronic configurations 
3s$^2$3p$^6$4s$^2$ (Ca),
4d$^7$5s$^1$ (Ru),
2s$^2$2p$^4$ (O)
with a  $5\times5\times3$
Monkhorst-Pack $k$-point mesh \cite{Monkhorst/Pack:1976} 
and a 650~eV planewave cutoff.
},
and vary the on-site Coulomb repulsion from 0 to 2\,eV within the Ru $4d$ manifold.
This upper limit on $U$ is close to the constrained random-phase approximation 
and dynamical mean field theory (DMFT) values for 
the related ruthenate Ca$_2$RuO$_4$ \cite{pavarini:2017,SutterNatCom2017,Das_Chang}.
We then interpret these dependencies and assign them to transitions at $T_s=48$\,K and 
$T=30$\,K. 
Previous DFT calculations   have shown a strong 
interplay between SOI and Coulomb repulsion $U$ \cite{LiuGuo-Qiang2011, Zheting:2018}. 
For this reason, unless otherwise noted, 
we perform all calculations with SOI and Coulomb repulsion following the 
Dudarev formalism \cite{Dudarev/Sutton_et_al:1998}.
We also relax the atomic positions 
(forces $<$0.1\,meV\,\AA$^{-1}$) at the $\mathrm{DFT}+U+\mathrm{SOI}$ level using Gaussian smearing 
(20\,meV width), the experimental lattice  parameters at 8\,K \cite{yoshida2005}, and 
the AFM-$b$ magnetic order.
\begin{table}
\caption{\label{tab:bonds}Changes in local RuO$_6$ octahedral volume ($V$, \AA$^3$),  
$\mathrm{Ru-O}$ bond lengths ($\ell$, \AA), and $\mathrm{Ru-O-Ru}$ angles (average in-plane, out-of-plane) measured in degrees with Coulomb interaction $U$. 
The  Ru $8b$ site in $Bb2_1m$ \CRO\  splits into two $4a$ sites in $Pn2_1a$  as specified
in parentheses.
}
\begin{ruledtabular}
\begin{tabular}{lccc}
 & $U=0$\,eV & \multicolumn{2}{c}{$U=1.6$\,eV} \\
\cline{2-2}\cline{3-4}\\[-0.9em]
$V(\mathrm{RuO}_6)$ & 10.54	($8b$) & 	10.58 ($4a$)	&	10.44 ($4a$)	\\
\cline{2-2}\cline{3-3}\cline{4-4}\\[-0.9em]
$\ell(\mathrm{Ru-O})$%
&1.981, 1.983		&1.973, 1.977		&1.971, 1.979\\
&1.988, 1.992		&1.933, 2.002		&1.986, 1.990\\
&2.006, 2.009		&2.013, 2.015		&1.995, 1.997\\
\cline{2-2}\cline{3-4}\\[-0.9em]
$\angle(\mathrm{Ru-O-Ru})$%
& 149.71, 152.16 &  \multicolumn{2}{c}{150.03, 152.71}	 \\
\end{tabular}
\end{ruledtabular}
\end{table}

\section{Results and Discussion}

\subsection{Crystal Structure}

Upon varying $U$ and performing $\mathrm{DFT}+U+\mathrm{SOI}$ atomic relaxations, we 
find that \CRO\ undergoes a  structural transition for $U > 1.3$\,eV characterized by 
a change in translational symmetry. 
The $Bb2_1m$ structure with AFM-$b$ order is dynamically unstable and 
transforms to the primitive orthorhombic $Pn2_1a$ structure (\autoref{fig1}a). 
The polar $Pn2_1a$  structure is similar to the $Bb2_1m$ phase; however, the 
differential Ru-O bond distortions within the octahedra (\autoref{tab:bonds}), which 
resemble breathing-like distortions, remove the  $B$-centering operation and permit a charge density wave (CDW). 
The local distortions in $Pn2_1a$ tile in a rock-salt-like arrangement of 
small (S) and large (L) RuO$_6$ octahedra (\autoref{fig1}a). \dpg{For this reason, while the $Bb2_1m$ phase is described using two formula units, the $Pn2_1a$  structure  is described using four formula units. Also, as indicated by the space group symmetry, the $b$-glide and mirror plane in the $Bb2_1m$ structure transform to $n$- and $a$-glide operations in the  $Pn2_1a$ phase, respectively.
This subtle change in crystallography gives weak Bragg reflections of the type $(0,k,l)$ with $k,l$ odd for $Pn2_1a$ that are absent in $Bb2_1m$.
Importantly, convergent-beam electron diffraction (CBED) experiments reported in Ref.~\onlinecite{yoshida2005} are performed in the $(h,k,0)$ scattering plane and not $(0,k,l)$, which prohibits the reported experiment geometry from differentiating between these two space groups.}

\autoref{fig1}b shows the octahedral volume ($V$) expansion/contraction are 
modulated by the degree of Coulomb repulsion, which we quantify as $B(\mathrm{RuO}_6)$ in 
\autoref{fig1}c using the percent difference in volumes 
of two adjacent RuO$_6$. 
As $U$ increases in the $Pn2_1a$ phase, the average RuO$_6$ octahedra volume decreases 
following the  trend exhibited by the $Bb2_1m$ structure. 
The average in-plane and out-of-plane $\mathrm{Ru-O-Ru}$ angles (\autoref{tab:bonds} and Ref.\ \onlinecite{suppmat}) 
exhibit a small increase with $U$, which is in remarkable agreement with the measured temperature dependence below  
$T_s$ \cite{yoshida2005}.
The Ca displacements that lift inversion symmetry undergo minimal changes owing 
to weak coupling of these displacements to the low-energy states on the Fermi surface 
\cite{puggioni:2013}.
Importantly, we are unable to stabilize the $Pn2_1a$ phase with  AFM-$a$ magnetic ordering, confirming 
the  cooperation among Coulomb repulsion, magnetic ordering,  and structural degrees of freedom 
in \CRO.

Coincident with the symmetry reduction, 
we find disproportionation of the local Ru spin moments, 
suggesting a spin density wave (SDW) occurs along with the CDW. 
The small (large) RuO$_6$ octahedra exhibits a %
larger (smaller) magnetic moment compared to the adjacent larger (smaller) 
octahedra (\autoref{fig1}d). 
This difference increases with increasing $U$, reaching  $\approx0.2$\,$\mu_B$ for $U=2$\,eV. 
\dpg{Owing to the spin disproportionation, the magnetic space group changes from $P_Cna2_1$ (No.~33.154, crystal class $mm21^\prime$  \cite{Lovesey:2019}) to  $Pna2_1$ (No.~33.144), crystal class $mm2$)  \footnote{We use the Belov-Neronova-Smirnova (BNS) setting of magnetic space groups, see Bilbao Crystallographic server, \href{http://www.cryst.ehu.es}{http://www.cryst.ehu.es}.}}.
Unlike the breathing distortion $B(\mathrm{RuO}_6)$,  the average magnetic moment 
increases with increasing Coulomb repulsion indicating overall localization of the itinerant 
holes within the nominal $t_{2g}$ manifold of Ru.
The coordinations of the small (large) RuO$_6$ octahedra \cite{suppmat} suggest 
weak G-type orbital order with the $d_{xy}$ orbital lower (higher) in energy for the large (small) RuO$_6$ octahedra. 
Moreover, we find a weak orbital moment of $\approx8\times10^{-3}$~$\mu_B$.

Spin-orbit interaction plays a key role in stabilizing the $Pn2_1a$ phase in the presence
of the aforementioned degrees-of-freedom. 
Upon excluding SOI but allowing full atomic relaxations, we find the $Pn2_1a$ phase 
remains more stable than the $Bb2_1m$ phase for $U > 1.1$~eV (\autoref{fig1}e). 
For $U>1.4$\,eV, the total energy difference of the $Pn2_1a$ structures with and without SOI is 
$\approx0.1$\,eV per formula unit (f.u.).
However, the spontaneous relaxation  of $Bb2_1m \rightarrow Pn2_1a$ does 
not occur---both phases are metastable and separated by an energy barrier, despite 
the one-dimensional $Y_2$ order parameter for the symmetry breaking displacements 
permitting the transition to be continuous.
This can be understood as follows: The inclusion of SOI in the calculations reduces 
the magnetic symmetry from 
$C_{1h}$ to $C_1$ (loss of inversion) allowing energy-lowering displacements due to the broken symmetry. To confirm this behavior, we performed a structural relaxation with $U=1.6$\,eV excluding SOI starting from the $Bb2_1m$ phase and removing all symmetries from the calculation.  We found that the $Bb2_1m$ phase relaxes into the triclinic $P1$ space group, which exhibits the same type of breathing distortion present in the $Pn2_1a$ phase, but is slightly higher in energy than the $Pn2_1a$ ground state \cite{suppmat}.
For these reasons, an accurate atomic ground state description of \CRO\ requires treating both 
SOI and Coulomb repulsion. 
 
\begin{figure}[]
\centering
\includegraphics[width=\columnwidth]{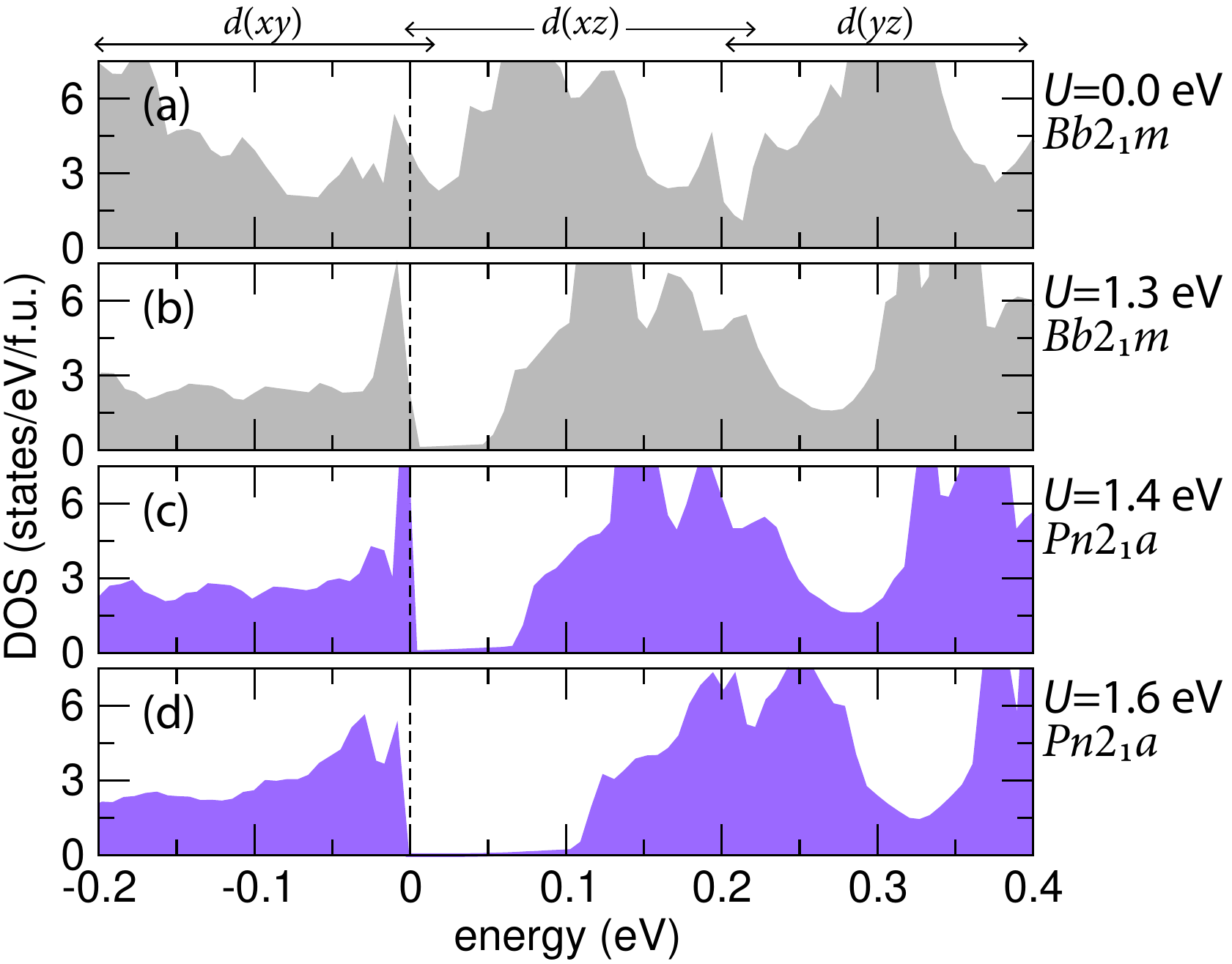}\vspace{-0pt}
  \caption{$\mathrm{DFT}+U+\mathrm{SOI}$ calculated total 
  density-of-states (DOS)  for \CRO. 
  $E_F=0$\,eV (broken line).
  }
 \label{dos_vs_U}
\end{figure}

\begin{figure*}
\includegraphics[width=0.95\textwidth]{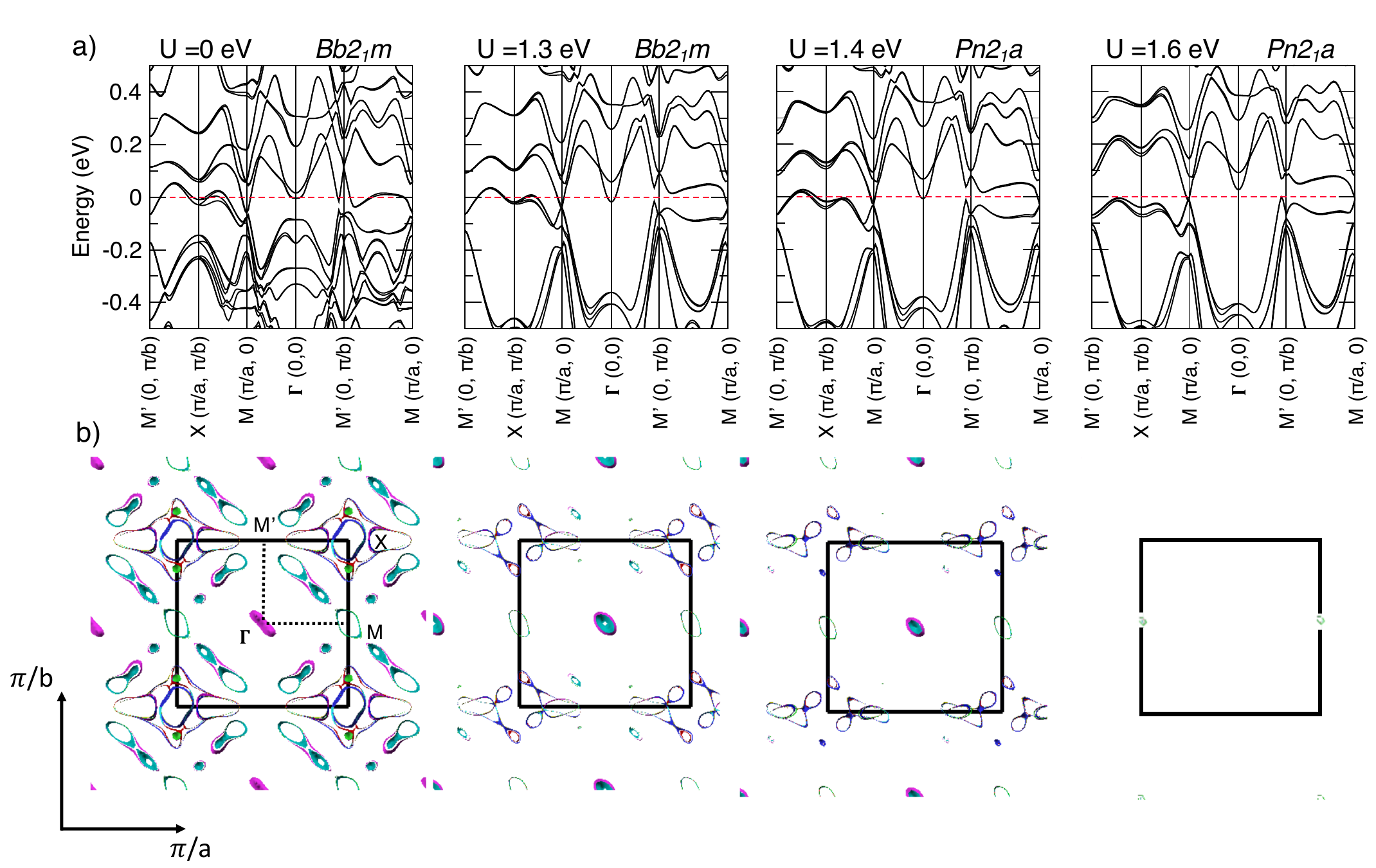}\vspace{-10pt}
\caption{$\mathrm{DFT}+U+\mathrm{SOI}$ calculated (a) electronic band structures  and (b) Fermi surfaces in the $k_z=0$ plane. The Fermi level is at 0~eV (broken line).
}
\label{bnd_vs_U}
\end{figure*} 
 


\subsection{Electronic Structure}

\autoref{dos_vs_U} shows the total density of states (DOS) of \CRO\ as a function of the Coulomb-repulsion strength. 
The low-energy electronic structure consists of Ru $4d$ states, nominally with low-spin 
Ru$^{4+}$ in a $d^2(xy)d^1(yz)d^1(xz)$ configuration hybridized with O $2p$ states  \cite{shiming:2018}. 
For $U=0$~eV, we find that the valence band is primarily $d(xy)$ with nearly full occupancy while 
the conduction band consists of primarily unoccupied $d(xz)$-derived states, spanning from the Fermi level ($E_F$)  
to $\approx0.2$\,eV followed by the $d(yz)$ orbitals located higher in energy ($\approx0.2\mathrm{-}0.4$\,eV).
The orbital hierarchy is a consequence of the crystal-field-split states and ratio of the lattice constants 
$b/a>1$; furthermore, the layered structure enhances the bandwidth of the $d(xy)$-derived bands relative to 
the $d(xz)/d(yz)$-derived states.
As a consequence, the band structure is nearly semimetallic with multiple bands 
crossing $E_F$ and exhibits quasi-2D character (\autoref{bnd_vs_U}):
The fine structure in the DOS below $E_F$ arises from the Ru multiplicity for the $d(xy)$-derived 
states, which are degenerate  at \dpg{M$^\prime(0,\pi/b)$} but split off from each other at the zone boundary 
\dpg{X$(\pi/a,\pi/b)$}, as shown in \autoref{bnd_vs_U}, and the extension of the $d(xz)$-derived bands below 
$E_F$. These features evolve with $U$ as we discuss  below.
SOI also lifts spin degeneracies along particular paths in the Brillouin zone 
due to Rashba and Dresselhaus interactions, \emph{e.g.}, for the aforementioned $d(xy)$-derived states 
spanning $E_F$ along the \dpg{X$(\pi/a,\pi/b)$-M$(\pi/a,0)$} trajectory.
Independent of the crystal structure, a metal-semimetal transition occurs as 
a pseudogap opens at $E_F$ for $U>0$~eV.
The width of the pseudogap depends on $U$  consistent with the temperature dependence of 
the pseudogap obtained from $ab$-plane optical conductivity spectra \cite{Leejs:2007}.
Raman scattering performed on single crystals of \CRO\ indicate  
a pseudo-gap of $\approx96$\,meV appears at $T_s$ \cite{PhysRevB.60.R6980}, suggesting that 
any model of the low-temperature phases of \CRO\ should include Coulomb 
interactions (and SOI).
The pseudogap in the $Bb2_1m$ phase arises from the $U$ interaction favoring 
preferential orbital occupancy and a redistribution of charge 
among the $d(xy)$ and $d(xz)$  orbitals along with small bandwidth narrowing. 
Indeed, while the $d(xy)$-derived bands located at \dpg{X$(\pi/a,\pi/b)$} shift to lower energy with $U$, 
the $d(xz)$-derived bands located at $\Gamma$ (and $d(yz)$-derived states) shift to higher energy 
(\autoref{dos_vs_U} and Ref.\ \onlinecite{suppmat}).
The SOI interaction with $U\ne0$ increases the energy separation among fractionally 
occupied states relative to those are occupied \cite{Zhang2018}, because 
the $d(xy)$, $d(xz)$, and $d(yz)$ states are already split by the $C_1$ 
crystal field (\autoref{tab:bonds}) and the $d(xy)$-like band is close to integer filling.
As a result, we observe a van Hove singularity (vHs) near $E_F$ (\autoref{dos_vs_U}), 
which is sensitive to small changes in $U$ and hence its position and access to it 
should be temperature and sample purity dependent.
This places \CRO\ proximate to an electronic instability in the $Bb2_1m$ phase, and 
susceptible to a Fermi surface reconstruction.
For this reason, we believe  the temperature-dependent thermopower data 
in Ref.~\onlinecite{Xing:2018} should be interpreted with caution since the sign of 
thermopower cannot be linked directly to the type of carrier when a vHs is
within $\approx$10\,meV of $E_F$ \cite{McIntosh:1996}.
To understand the origin of the vHs, we next examine in further detail 
the band dispersions, focusing on the $k_z=0$ 2D slice of the Brillouin zone (\autoref{bnd_vs_U}) because 
dispersions along $k_z$ are weak \cite{LiuGuo-Qiang2011}.
For $U=0$\,eV, the Ru bands give rise to quasi-2D hole and electron  pockets 
along \dpg{M$^\prime(0,\pi/b)\rightarrow$X$(\pi/a,\pi/b)$} with the electron pockets centered on \dpg{X$(\pi/a,\pi/b)$}. 
We find an electron-like knob at the $\Gamma$ point along with 
small quasi-2D electron pockets about $\Gamma$ and \dpg{M$(\pi/a,0)$}. 
These features are clearly discernible in the  FS (\autoref{bnd_vs_U}b), 
which presents small degenerate electron pockets centered at 
\dpg{X$(\pi/a,\pi/b)$} and \dpg{M$(\pi/a,0)$},  degenerate hole pockets bordering 
the electron pocket at \dpg{X$(\pi/a,\pi/b)$}, and a feature which almost connects 
a linear trajectory from \dpg{M$(\pi/a,0)$ to M$^\prime(0,\pi/b)$}. 
The changes in carrier type in momentum space are due to a saddle point near 
$E_F$ along the \dpg{M$^\prime(0,\pi/b)$--X$(\pi/a,\pi/b)$--M$(\pi/a,0)$} trajectory.
In addition, nearly flat bands cross $E_F$ along the \dpg{M$^\prime(0,\pi/b)$--M$(\pi/a,0)$} line; 
beginning with $U=1.3$~eV, a gap opens along this trajectory and persists (\autoref{bnd_vs_U}).
The orbital-energy degeneracies within 25\,meV of $E_F$ at \dpg{X$(\pi/a,\pi/b)$} are 
responsible for the identified vHs in the DOS (\autoref{dos_vs_U}). 
\autoref{bnd_vs_U} also allows us to deduce that the electron pockets centered near  
\dpg{M$(\pi/a,0)$} originate from a Dirac-like feature, located at $\approx70$~meV below $E_F$, 
which has been recently detected by ARPES experiments \cite{ARPES2019}. 
These features shift away from \dpg{M$(\pi/a,0)$} owing to SOI and the broken 
parity and time reversal symmetries in \CRO.

\begin{figure*}[]
\begin{center}
\includegraphics[width=0.95\textwidth]{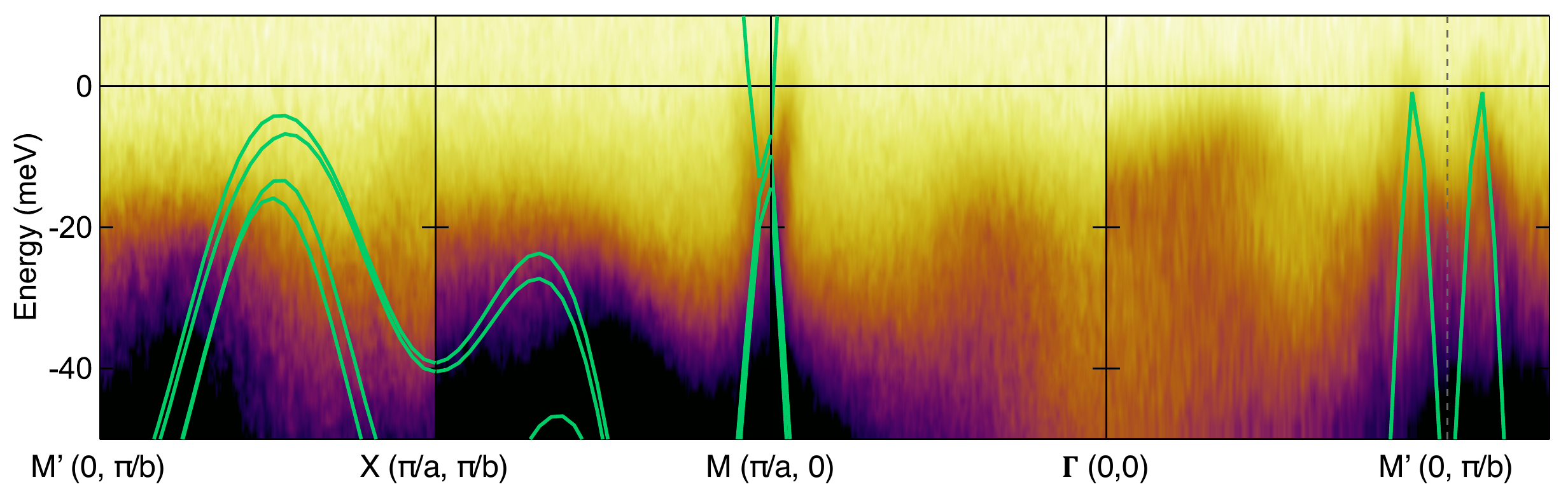}\vspace{-10pt}
  \caption{Comparison of the $\mathrm{DFT}+U+\mathrm{SOI}$ ($U=1.6$\,eV) 
  band dispersions with those from Ref.~\onlinecite{ARPES2019} acquired at 16\,K.}
 \label{DFTvsARPES}
 \end{center}
\end{figure*} 

The change in crystal structure also affects the vHs and low-energy 
electronic structure for $U>1.3$\,eV.
\autoref{dos_vs_U}c shows that the vHs splits just below $E_F$, because the 
change in the local Ru crystal field from the breathing distortions lifts the 
degeneracy of the single-particle $d(xy)$-derived eigen-energies from different RuO$_6$ octahedra 
[see \autoref{bnd_vs_U}a at \dpg{X$(\pi/a,\pi/b)$} orbitals].
Upon increasing $U$, these states shift to lower energy while the electron pocket 
derived from the $d(xz)$ orbital is fully de-occupied at $\Gamma$.
The bandwidth is further reduced with increasing $U$, which also enhances the SOI 
spin splittings. %
Cooperatively, these effects shift the vHs to lower energy (\autoref{dos_vs_U}d) and 
reduce the FS to comprise small electron pockets at \dpg{M$(\pi/a,0)$} 
proximate to Weyl nodes (\autoref{bnd_vs_U}).
These findings confirm the importance of SOI and Coulomb repulsion in the treatment of the low  temperature phase of \CRO\ \cite{LiuGuo-Qiang2011}.  
For $0<U<1.6$\,eV,  the area of the Brillouin zone occupied by the FS is reduced.
The volume of the electron pockets  centered about \dpg{M$(\pi/a,0)$}  decreases and 
the features along the  \dpg{M$^\prime(0,\pi/b)$--M$(\pi/a,0)$} line and at the $\Gamma$ point vanish.
Due to the shift of the vHs, the electron pockets centered on  \dpg{X$(\pi/a,\pi/b)$} also disappear while the nearby hole pockets survive.  
For $U=1.6$~eV, only the electron pockets about \dpg{M$(\pi/a,0)$} survive---all bands are gapped with the exception of the Dirac-like structure that shifts to $\sim15$~meV below $E_F$. 
In addition, we find that a full charge gap ($\approx30$\,meV) appears for $U=1.8$\,eV with SOI.
This critical value of $U$ for the metal-insulator transition in \CRO\ is well below $U=3$~eV  reported in Ref.~\cite{LiuGuo-Qiang2011}, which did not allow for local changes in 
the atomic structure at the $\mathrm{DFT}+U+\mathrm{SOI}$ level. 
Importantly, no insulating state is reported for \CRO; hence, $U=1.8$\,eV 
is the upper bound on the interaction strength within our model.

\subsection{Phase Transition Assignments}
%
The addition of Coulomb repulsion within our DFT calculations 
has three major effects: $(i)$ it shifts the position of the vHs to lower energy, 
$(ii)$ reduces the overlap between the top of the valence band and the bottom of the conduction band, and 
$(iii)$ shifts the Dirac-like feature at \dpg{M$(\pi/a,0)$} to higher energy. 
Therefore, it is natural to ask what is the correct value for $U$ and how can it be used to model the temperature-dependent fermiology. To this purpose, we now compare our calculated electronic band structures 
as a function of $U+\mathrm{SOI}$ with ARPES measured dispersions at 16\,K \cite{ARPES2019}.
We find best agreement for $U=1.6$\,eV (\autoref{DFTvsARPES}). 
In particular, the Dirac-like feature at $(\pi/a,0)$ and the electronic band structure along  
$\Gamma$-\dpg{M$^\prime(0,\pi/b)$} are well captured. 
Note that for $U>1.3$\,eV, we predict a change in 
symmetry ($Bb2_1m \rightarrow Pn2_1a$) for \CRO. 
Using this as an approximate bound on the ground state $U$ value for \CRO, 
we assign  $T=30$\,K to be a CDW/SDW transition.

Furthermore, reasonable values of $U$ to model intermediate temperatures $T_\mathrm{CDW}=30<T<T_s=48$\,K 
then should span values of $0.5<U<1.6$\,eV. 
This is justified by the increase of the pseudogap with temperature revealed by 
optical spectroscopy \cite{Leejs:2007}.
For this range of Coulomb interactions, we find that weak static correlations 
with SOI and RuO$_6$ distortions drive a Liftshitz-like transition that 
reconstructs the Fermi surface and opens a pseudo-gap at $T_s$ due to filling of the $d(xy)$ bands 
preferentially to the $d(xz)$ band that is emptied, leaving behind a vHs and Weyl nodes just below 
$E_F$.
In this temperature range, it is plausible that the $Pn2_1a$ phase persists; further  experiments 
are required to conclusively resolve this conjecture.

\section{Conclusions}
Using DFT calculations combined with ARPES measurements, 
we reconciled existing experiments on the low temperature fermiology of \CRO. 
The cooperative interplay among Coulomb repulsion $U$, magnetic ordering, spin-orbit interactions, and structural degrees of freedom 
govern the temperature-dependent phase transitions. 
We assign a a Lifshitz-like transition to $T_s=48$\,K and 
a charge- (spin-) density wave transition to $T_\mathrm{CDW}=30$\,K, which were hypothesized from 
optical spectroscopy \cite{Leejs:2007}, along with broken translational symmetry ($Bb2_1m\rightarrow Pn2_1a$) at $T_\mathrm{CDW}$.
Last, we validated our temperature-dependent Coulomb interaction model with ARPES experiments at 16\,K  \cite{ARPES2019}, 
which indicated that the low-energy physics requires moderate Coulomb repulsion ($U=1.6$~eV) and the unique  $Pn2_1a$ symmetry. 
\dpg{We hope new structural assessments are performed at low temperature with advanced CBED or synchrotron scattering techniques to assess the predicted revised phase diagram.}
Our work points to the importance of treating multiple interactions on equal footing in correlated metals, 
which exhibit nontrivial fermions near the Fermi level and fractional orbital filling, as SOI and Coulombic interactions may conspire with lattice and spin degrees-of-freedom to produce unexpected electronic and structural transitions.

\begin{acknowledgments}
M.H.\ and J.C.\ thank M. H. Fischer for insightful discussions.
D.P.\ and J.M.R.\ acknowledge the Army Research Office under Grant No.\ 
W911NF-15-1-0017 for financial support and the DOD-HPCMP for computational resources. 
M.H., M.H.F., and J.C.\ acknowledge support from the Swiss National Science Foundation.
\end{acknowledgments}

\bibliography{puggioni}

\begin{thebibliography}{39}%
\makeatletter
\providecommand \@ifxundefined [1]{%
 \@ifx{#1\undefined}
}%
\providecommand \@ifnum [1]{%
 \ifnum #1\expandafter \@firstoftwo
 \else \expandafter \@secondoftwo
 \fi
}%
\providecommand \@ifx [1]{%
 \ifx #1\expandafter \@firstoftwo
 \else \expandafter \@secondoftwo
 \fi
}%
\providecommand \natexlab [1]{#1}%
\providecommand \enquote  [1]{``#1''}%
\providecommand \bibnamefont  [1]{#1}%
\providecommand \bibfnamefont [1]{#1}%
\providecommand \citenamefont [1]{#1}%
\providecommand \href@noop [0]{\@secondoftwo}%
\providecommand \href [0]{\begingroup \@sanitize@url \@href}%
\providecommand \@href[1]{\@@startlink{#1}\@@href}%
\providecommand \@@href[1]{\endgroup#1\@@endlink}%
\providecommand \@sanitize@url [0]{\catcode `\\12\catcode `\$12\catcode
  `\&12\catcode `\#12\catcode `\^12\catcode `\_12\catcode `\%12\relax}%
\providecommand \@@startlink[1]{}%
\providecommand \@@endlink[0]{}%
\providecommand \url  [0]{\begingroup\@sanitize@url \@url }%
\providecommand \@url [1]{\endgroup\@href {#1}{\urlprefix }}%
\providecommand \urlprefix  [0]{URL }%
\providecommand \Eprint [0]{\href }%
\providecommand \doibase [0]{http://dx.doi.org/}%
\providecommand \selectlanguage [0]{\@gobble}%
\providecommand \bibinfo  [0]{\@secondoftwo}%
\providecommand \bibfield  [0]{\@secondoftwo}%
\providecommand \translation [1]{[#1]}%
\providecommand \BibitemOpen [0]{}%
\providecommand \bibitemStop [0]{}%
\providecommand \bibitemNoStop [0]{.\EOS\space}%
\providecommand \EOS [0]{\spacefactor3000\relax}%
\providecommand \BibitemShut  [1]{\csname bibitem#1\endcsname}%
\let\auto@bib@innerbib\@empty
\bibitem [{\citenamefont {Cao}\ \emph {et~al.}(2003)\citenamefont {Cao},
  \citenamefont {Balicas}, \citenamefont {Xin}, \citenamefont {Dagotto},
  \citenamefont {Crow}, \citenamefont {Nelson},\ and\ \citenamefont
  {Agterberg}}]{Cao_Agterberg:2003}%
  \BibitemOpen
  \bibfield  {author} {\bibinfo {author} {\bibfnamefont {G.}~\bibnamefont
  {Cao}}, \bibinfo {author} {\bibfnamefont {L.}~\bibnamefont {Balicas}},
  \bibinfo {author} {\bibfnamefont {Y.}~\bibnamefont {Xin}}, \bibinfo {author}
  {\bibfnamefont {E.}~\bibnamefont {Dagotto}}, \bibinfo {author} {\bibfnamefont
  {J.~E.}\ \bibnamefont {Crow}}, \bibinfo {author} {\bibfnamefont {C.~S.}\
  \bibnamefont {Nelson}}, \ and\ \bibinfo {author} {\bibfnamefont {D.~F.}\
  \bibnamefont {Agterberg}},\ }\href {\doibase 10.1103/PhysRevB.67.060406}
  {\bibfield  {journal} {\bibinfo  {journal} {Phys. Rev. B}\ }\textbf {\bibinfo
  {volume} {67}},\ \bibinfo {pages} {060406} (\bibinfo {year}
  {2003})}\BibitemShut {NoStop}%
\bibitem [{\citenamefont {Lin}\ \emph {et~al.}(2005)\citenamefont {Lin},
  \citenamefont {Zhou}, \citenamefont {Durairaj}, \citenamefont {Schlottmann},\
  and\ \citenamefont {Cao}}]{Lin_Cao:2005}%
  \BibitemOpen
  \bibfield  {author} {\bibinfo {author} {\bibfnamefont {X.~N.}\ \bibnamefont
  {Lin}}, \bibinfo {author} {\bibfnamefont {Z.~X.}\ \bibnamefont {Zhou}},
  \bibinfo {author} {\bibfnamefont {V.}~\bibnamefont {Durairaj}}, \bibinfo
  {author} {\bibfnamefont {P.}~\bibnamefont {Schlottmann}}, \ and\ \bibinfo
  {author} {\bibfnamefont {G.}~\bibnamefont {Cao}},\ }\href {\doibase
  10.1103/PhysRevLett.95.017203} {\bibfield  {journal} {\bibinfo  {journal}
  {Phys. Rev. Lett.}\ }\textbf {\bibinfo {volume} {95}},\ \bibinfo {pages}
  {017203} (\bibinfo {year} {2005})}\BibitemShut {NoStop}%
\bibitem [{\citenamefont {Cao}\ \emph {et~al.}(2004)\citenamefont {Cao},
  \citenamefont {Lin}, \citenamefont {Balicas}, \citenamefont {Chikara},
  \citenamefont {Crow},\ and\ \citenamefont {Schlottmann}}]{Cao_2004}%
  \BibitemOpen
  \bibfield  {author} {\bibinfo {author} {\bibfnamefont {G.}~\bibnamefont
  {Cao}}, \bibinfo {author} {\bibfnamefont {X.~N.}\ \bibnamefont {Lin}},
  \bibinfo {author} {\bibfnamefont {L.}~\bibnamefont {Balicas}}, \bibinfo
  {author} {\bibfnamefont {S.}~\bibnamefont {Chikara}}, \bibinfo {author}
  {\bibfnamefont {J.~E.}\ \bibnamefont {Crow}}, \ and\ \bibinfo {author}
  {\bibfnamefont {P.}~\bibnamefont {Schlottmann}},\ }\href {\doibase
  10.1088/1367-2630/6/1/159} {\bibfield  {journal} {\bibinfo  {journal} {New
  Journal of Physics}\ }\textbf {\bibinfo {volume} {6}},\ \bibinfo {pages}
  {159} (\bibinfo {year} {2004})}\BibitemShut {NoStop}%
\bibitem [{\citenamefont {Lei}\ \emph {et~al.}(2018)\citenamefont {Lei},
  \citenamefont {Gu}, \citenamefont {Puggioni}, \citenamefont {Stone},
  \citenamefont {Peng}, \citenamefont {Ge}, \citenamefont {Wang}, \citenamefont
  {Wang}, \citenamefont {Yuan}, \citenamefont {Wang}, \citenamefont {Mao},
  \citenamefont {Rondinelli},\ and\ \citenamefont {Gopalan}}]{shiming:2018}%
  \BibitemOpen
  \bibfield  {author} {\bibinfo {author} {\bibfnamefont {S.}~\bibnamefont
  {Lei}}, \bibinfo {author} {\bibfnamefont {M.}~\bibnamefont {Gu}}, \bibinfo
  {author} {\bibfnamefont {D.}~\bibnamefont {Puggioni}}, \bibinfo {author}
  {\bibfnamefont {G.}~\bibnamefont {Stone}}, \bibinfo {author} {\bibfnamefont
  {J.}~\bibnamefont {Peng}}, \bibinfo {author} {\bibfnamefont {J.}~\bibnamefont
  {Ge}}, \bibinfo {author} {\bibfnamefont {Y.}~\bibnamefont {Wang}}, \bibinfo
  {author} {\bibfnamefont {B.}~\bibnamefont {Wang}}, \bibinfo {author}
  {\bibfnamefont {Y.}~\bibnamefont {Yuan}}, \bibinfo {author} {\bibfnamefont
  {K.}~\bibnamefont {Wang}}, \bibinfo {author} {\bibfnamefont {Z.}~\bibnamefont
  {Mao}}, \bibinfo {author} {\bibfnamefont {J.~M.}\ \bibnamefont {Rondinelli}},
  \ and\ \bibinfo {author} {\bibfnamefont {V.}~\bibnamefont {Gopalan}},\ }\href
  {https://doi.org/10.1021/acs.nanolett.8b00633} {\bibfield  {journal}
  {\bibinfo  {journal} {Nano Letters}\ }\textbf {\bibinfo {volume} {18}},\
  \bibinfo {pages} {3088} (\bibinfo {year} {2018})}\BibitemShut {NoStop}%
\bibitem [{\citenamefont {Stone}\ \emph {et~al.}(2019)\citenamefont {Stone},
  \citenamefont {Puggioni}, \citenamefont {Lei}, \citenamefont {Gu},
  \citenamefont {Wang}, \citenamefont {Wang}, \citenamefont {Ge}, \citenamefont
  {Lu}, \citenamefont {Mao}, \citenamefont {Rondinelli},\ and\ \citenamefont
  {Gopalan}}]{Stone:2019}%
  \BibitemOpen
  \bibfield  {author} {\bibinfo {author} {\bibfnamefont {G.}~\bibnamefont
  {Stone}}, \bibinfo {author} {\bibfnamefont {D.}~\bibnamefont {Puggioni}},
  \bibinfo {author} {\bibfnamefont {S.}~\bibnamefont {Lei}}, \bibinfo {author}
  {\bibfnamefont {M.}~\bibnamefont {Gu}}, \bibinfo {author} {\bibfnamefont
  {K.}~\bibnamefont {Wang}}, \bibinfo {author} {\bibfnamefont {Y.}~\bibnamefont
  {Wang}}, \bibinfo {author} {\bibfnamefont {J.}~\bibnamefont {Ge}}, \bibinfo
  {author} {\bibfnamefont {X.-Z.}\ \bibnamefont {Lu}}, \bibinfo {author}
  {\bibfnamefont {Z.}~\bibnamefont {Mao}}, \bibinfo {author} {\bibfnamefont
  {J.~M.}\ \bibnamefont {Rondinelli}}, \ and\ \bibinfo {author} {\bibfnamefont
  {V.}~\bibnamefont {Gopalan}},\ }\href {\doibase 10.1103/PhysRevB.99.014105}
  {\bibfield  {journal} {\bibinfo  {journal} {Phys. Rev. B}\ }\textbf {\bibinfo
  {volume} {99}},\ \bibinfo {pages} {014105} (\bibinfo {year}
  {2019})}\BibitemShut {NoStop}%
\bibitem [{\citenamefont {Christy}(1995)}]{Christy1995}%
  \BibitemOpen
  \bibfield  {author} {\bibinfo {author} {\bibfnamefont {A.~G.}\ \bibnamefont
  {Christy}},\ }\href {\doibase 10.1107/s0108768195001728} {\bibfield
  {journal} {\bibinfo  {journal} {Acta Cryst.\ B}\ }\textbf {\bibinfo {volume}
  {51}},\ \bibinfo {pages} {753} (\bibinfo {year} {1995})}\BibitemShut
  {NoStop}%
\bibitem [{\citenamefont {Yoshida}\ \emph {et~al.}(2005)\citenamefont
  {Yoshida}, \citenamefont {Ikeda}, \citenamefont {Matsuhata}, \citenamefont
  {Shirakawa}, \citenamefont {Lee},\ and\ \citenamefont
  {Katano}}]{yoshida2005}%
  \BibitemOpen
  \bibfield  {author} {\bibinfo {author} {\bibfnamefont {Y.}~\bibnamefont
  {Yoshida}}, \bibinfo {author} {\bibfnamefont {S.-I.}\ \bibnamefont {Ikeda}},
  \bibinfo {author} {\bibfnamefont {H.}~\bibnamefont {Matsuhata}}, \bibinfo
  {author} {\bibfnamefont {N.}~\bibnamefont {Shirakawa}}, \bibinfo {author}
  {\bibfnamefont {C.~H.}\ \bibnamefont {Lee}}, \ and\ \bibinfo {author}
  {\bibfnamefont {S.}~\bibnamefont {Katano}},\ }\href {\doibase
  10.1103/PhysRevB.72.054412} {\bibfield  {journal} {\bibinfo  {journal} {Phys.
  Rev. B}\ }\textbf {\bibinfo {volume} {72}},\ \bibinfo {pages} {054412}
  (\bibinfo {year} {2005})}\BibitemShut {NoStop}%
\bibitem [{\citenamefont {Basov}\ \emph {et~al.}(2017)\citenamefont {Basov},
  \citenamefont {Averitt},\ and\ \citenamefont {Hsieh}}]{Basov2017}%
  \BibitemOpen
  \bibfield  {author} {\bibinfo {author} {\bibfnamefont {D.~N.}\ \bibnamefont
  {Basov}}, \bibinfo {author} {\bibfnamefont {R.~D.}\ \bibnamefont {Averitt}},
  \ and\ \bibinfo {author} {\bibfnamefont {D.}~\bibnamefont {Hsieh}},\ }\href
  {\doibase 10.1038/nmat5017} {\bibfield  {journal} {\bibinfo  {journal}
  {Nature Materials}\ }\textbf {\bibinfo {volume} {16}},\ \bibinfo {pages}
  {1077} (\bibinfo {year} {2017})}\BibitemShut {NoStop}%
\bibitem [{\citenamefont {Baumberger}\ \emph {et~al.}(2006)\citenamefont
  {Baumberger}, \citenamefont {Ingle}, \citenamefont {Kikugawa}, \citenamefont
  {Hossain}, \citenamefont {Meevasana}, \citenamefont {Perry}, \citenamefont
  {Shen}, \citenamefont {Lu}, \citenamefont {Damascelli}, \citenamefont {Rost},
  \citenamefont {Mackenzie}, \citenamefont {Hussain},\ and\ \citenamefont
  {Shen}}]{Baumberger:2006}%
  \BibitemOpen
  \bibfield  {author} {\bibinfo {author} {\bibfnamefont {F.}~\bibnamefont
  {Baumberger}}, \bibinfo {author} {\bibfnamefont {N.~J.~C.}\ \bibnamefont
  {Ingle}}, \bibinfo {author} {\bibfnamefont {N.}~\bibnamefont {Kikugawa}},
  \bibinfo {author} {\bibfnamefont {M.~A.}\ \bibnamefont {Hossain}}, \bibinfo
  {author} {\bibfnamefont {W.}~\bibnamefont {Meevasana}}, \bibinfo {author}
  {\bibfnamefont {R.~S.}\ \bibnamefont {Perry}}, \bibinfo {author}
  {\bibfnamefont {K.~M.}\ \bibnamefont {Shen}}, \bibinfo {author}
  {\bibfnamefont {D.~H.}\ \bibnamefont {Lu}}, \bibinfo {author} {\bibfnamefont
  {A.}~\bibnamefont {Damascelli}}, \bibinfo {author} {\bibfnamefont
  {A.}~\bibnamefont {Rost}}, \bibinfo {author} {\bibfnamefont {A.~P.}\
  \bibnamefont {Mackenzie}}, \bibinfo {author} {\bibfnamefont {Z.}~\bibnamefont
  {Hussain}}, \ and\ \bibinfo {author} {\bibfnamefont {Z.-X.}\ \bibnamefont
  {Shen}},\ }\href {\doibase 10.1103/PhysRevLett.96.107601} {\bibfield
  {journal} {\bibinfo  {journal} {Phys. Rev. Lett.}\ }\textbf {\bibinfo
  {volume} {96}},\ \bibinfo {pages} {107601} (\bibinfo {year}
  {2006})}\BibitemShut {NoStop}%
\bibitem [{\citenamefont {Markovi\'c}\ \emph {et~al.}(2020)\citenamefont
  {Markovi\'c}, \citenamefont {Watson}, \citenamefont {Clark}, \citenamefont
  {Mazzola}, \citenamefont {Morales}, \citenamefont {Hooley}, \citenamefont
  {Rosner}, \citenamefont {Polley}, \citenamefont {Balasubramanian},
  \citenamefont {Mukherjee}, \citenamefont {Kikugawa}, \citenamefont {Sokolov},
  \citenamefont {Mackenzie},\ and\ \citenamefont {King}}]{spinreorientation}%
  \BibitemOpen
  \bibfield  {author} {\bibinfo {author} {\bibfnamefont {I.}~\bibnamefont
  {Markovi\'c}}, \bibinfo {author} {\bibfnamefont {M.~D.}\ \bibnamefont
  {Watson}}, \bibinfo {author} {\bibfnamefont {O.~J.}\ \bibnamefont {Clark}},
  \bibinfo {author} {\bibfnamefont {F.}~\bibnamefont {Mazzola}}, \bibinfo
  {author} {\bibfnamefont {E.~A.}\ \bibnamefont {Morales}}, \bibinfo {author}
  {\bibfnamefont {C.~A.}\ \bibnamefont {Hooley}}, \bibinfo {author}
  {\bibfnamefont {H.}~\bibnamefont {Rosner}}, \bibinfo {author} {\bibfnamefont
  {C.~M.}\ \bibnamefont {Polley}}, \bibinfo {author} {\bibfnamefont
  {T.}~\bibnamefont {Balasubramanian}}, \bibinfo {author} {\bibfnamefont
  {S.}~\bibnamefont {Mukherjee}}, \bibinfo {author} {\bibfnamefont
  {N.}~\bibnamefont {Kikugawa}}, \bibinfo {author} {\bibfnamefont {D.~A.}\
  \bibnamefont {Sokolov}}, \bibinfo {author} {\bibfnamefont {A.~P.}\
  \bibnamefont {Mackenzie}}, \ and\ \bibinfo {author} {\bibfnamefont
  {P.~D.~C.}\ \bibnamefont {King}},\ }\href@noop {} {\enquote {\bibinfo {title}
  {{Electronically driven spin-reorientation transition of the correlated polar
  metal Ca$_3$Ru$_2$O$_7$}},}\ } (\bibinfo {year} {2020}),\ \Eprint
  {http://arxiv.org/abs/2001.09499} {arXiv:2001.09499 [cond-mat.str-el]}
  \BibitemShut {NoStop}%
\bibitem [{\citenamefont {Yuan}\ \emph {et~al.}(2019)\citenamefont {Yuan},
  \citenamefont {Kissin}, \citenamefont {Puggioni}, \citenamefont {Cremin},
  \citenamefont {Lei}, \citenamefont {Wang}, \citenamefont {Mao}, \citenamefont
  {Rondinelli}, \citenamefont {Averitt},\ and\ \citenamefont
  {Gopalan}}]{Yuan:2019}%
  \BibitemOpen
  \bibfield  {author} {\bibinfo {author} {\bibfnamefont {Y.}~\bibnamefont
  {Yuan}}, \bibinfo {author} {\bibfnamefont {P.}~\bibnamefont {Kissin}},
  \bibinfo {author} {\bibfnamefont {D.}~\bibnamefont {Puggioni}}, \bibinfo
  {author} {\bibfnamefont {K.}~\bibnamefont {Cremin}}, \bibinfo {author}
  {\bibfnamefont {S.}~\bibnamefont {Lei}}, \bibinfo {author} {\bibfnamefont
  {Y.}~\bibnamefont {Wang}}, \bibinfo {author} {\bibfnamefont {Z.}~\bibnamefont
  {Mao}}, \bibinfo {author} {\bibfnamefont {J.~M.}\ \bibnamefont {Rondinelli}},
  \bibinfo {author} {\bibfnamefont {R.~D.}\ \bibnamefont {Averitt}}, \ and\
  \bibinfo {author} {\bibfnamefont {V.}~\bibnamefont {Gopalan}},\ }\href
  {\doibase 10.1103/PhysRevB.99.155111} {\bibfield  {journal} {\bibinfo
  {journal} {Phys. Rev. B}\ }\textbf {\bibinfo {volume} {99}},\ \bibinfo
  {pages} {155111} (\bibinfo {year} {2019})}\BibitemShut {NoStop}%
\bibitem [{\citenamefont {Horio}\ \emph {et~al.}(2019)\citenamefont {Horio},
  \citenamefont {Wang}, \citenamefont {Granata}, \citenamefont {Kramer},
  \citenamefont {Sassa}, \citenamefont {J\"ohr}, \citenamefont {Sutter},
  \citenamefont {Bold}, \citenamefont {Das}, \citenamefont {Xu}, \citenamefont
  {Frison}, \citenamefont {Fittipaldi}, \citenamefont {Kim}, \citenamefont
  {Cacho}, \citenamefont {Rault}, \citenamefont {F\'evre}, \citenamefont
  {Bertran}, \citenamefont {Plumb}, \citenamefont {Shi}, \citenamefont
  {Vecchione}, \citenamefont {Fischer},\ and\ \citenamefont
  {Chang}}]{ARPES2019}%
  \BibitemOpen
  \bibfield  {author} {\bibinfo {author} {\bibfnamefont {M.}~\bibnamefont
  {Horio}}, \bibinfo {author} {\bibfnamefont {Q.}~\bibnamefont {Wang}},
  \bibinfo {author} {\bibfnamefont {V.}~\bibnamefont {Granata}}, \bibinfo
  {author} {\bibfnamefont {K.~P.}\ \bibnamefont {Kramer}}, \bibinfo {author}
  {\bibfnamefont {Y.}~\bibnamefont {Sassa}}, \bibinfo {author} {\bibfnamefont
  {S.}~\bibnamefont {J\"ohr}}, \bibinfo {author} {\bibfnamefont
  {D.}~\bibnamefont {Sutter}}, \bibinfo {author} {\bibfnamefont
  {A.}~\bibnamefont {Bold}}, \bibinfo {author} {\bibfnamefont {L.}~\bibnamefont
  {Das}}, \bibinfo {author} {\bibfnamefont {Y.}~\bibnamefont {Xu}}, \bibinfo
  {author} {\bibfnamefont {R.}~\bibnamefont {Frison}}, \bibinfo {author}
  {\bibfnamefont {R.}~\bibnamefont {Fittipaldi}}, \bibinfo {author}
  {\bibfnamefont {T.~K.}\ \bibnamefont {Kim}}, \bibinfo {author} {\bibfnamefont
  {C.}~\bibnamefont {Cacho}}, \bibinfo {author} {\bibfnamefont {J.~E.}\
  \bibnamefont {Rault}}, \bibinfo {author} {\bibfnamefont {P.~L.}\ \bibnamefont
  {F\'evre}}, \bibinfo {author} {\bibfnamefont {F.}~\bibnamefont {Bertran}},
  \bibinfo {author} {\bibfnamefont {N.~C.}\ \bibnamefont {Plumb}}, \bibinfo
  {author} {\bibfnamefont {M.}~\bibnamefont {Shi}}, \bibinfo {author}
  {\bibfnamefont {A.}~\bibnamefont {Vecchione}}, \bibinfo {author}
  {\bibfnamefont {M.~H.}\ \bibnamefont {Fischer}}, \ and\ \bibinfo {author}
  {\bibfnamefont {J.}~\bibnamefont {Chang}},\ }\href@noop {} {\enquote
  {\bibinfo {title} {{Electron-driven $C_2$-symmetric Dirac semimetal uncovered
  in Ca$_3$Ru$_2$O$_7$}},}\ } (\bibinfo {year} {2019}),\ \Eprint
  {http://arxiv.org/abs/1911.12163} {arXiv:1911.12163 [cond-mat.str-el]}
  \BibitemShut {NoStop}%
\bibitem [{\citenamefont {Lee}\ \emph {et~al.}(2007)\citenamefont {Lee},
  \citenamefont {Moon}, \citenamefont {Yang}, \citenamefont {Yu}, \citenamefont
  {Schade}, \citenamefont {Yoshida}, \citenamefont {Ikeda},\ and\ \citenamefont
  {Noh}}]{Leejs:2007}%
  \BibitemOpen
  \bibfield  {author} {\bibinfo {author} {\bibfnamefont {J.~S.}\ \bibnamefont
  {Lee}}, \bibinfo {author} {\bibfnamefont {S.~J.}\ \bibnamefont {Moon}},
  \bibinfo {author} {\bibfnamefont {B.~J.}\ \bibnamefont {Yang}}, \bibinfo
  {author} {\bibfnamefont {J.}~\bibnamefont {Yu}}, \bibinfo {author}
  {\bibfnamefont {U.}~\bibnamefont {Schade}}, \bibinfo {author} {\bibfnamefont
  {Y.}~\bibnamefont {Yoshida}}, \bibinfo {author} {\bibfnamefont {S.-I.}\
  \bibnamefont {Ikeda}}, \ and\ \bibinfo {author} {\bibfnamefont {T.~W.}\
  \bibnamefont {Noh}},\ }\href {\doibase 10.1103/PhysRevLett.98.097403}
  {\bibfield  {journal} {\bibinfo  {journal} {Phys. Rev. Lett.}\ }\textbf
  {\bibinfo {volume} {98}},\ \bibinfo {pages} {097403} (\bibinfo {year}
  {2007})}\BibitemShut {NoStop}%
\bibitem [{\citenamefont {Xing}\ \emph {et~al.}(2018)\citenamefont {Xing},
  \citenamefont {Wen}, \citenamefont {Shen}, \citenamefont {He}, \citenamefont
  {Cai}, \citenamefont {Peng}, \citenamefont {Wang}, \citenamefont {Tian},
  \citenamefont {Xu}, \citenamefont {Ku}, \citenamefont {Mao},\ and\
  \citenamefont {Liu}}]{Xing:2018}%
  \BibitemOpen
  \bibfield  {author} {\bibinfo {author} {\bibfnamefont {H.}~\bibnamefont
  {Xing}}, \bibinfo {author} {\bibfnamefont {L.}~\bibnamefont {Wen}}, \bibinfo
  {author} {\bibfnamefont {C.}~\bibnamefont {Shen}}, \bibinfo {author}
  {\bibfnamefont {J.}~\bibnamefont {He}}, \bibinfo {author} {\bibfnamefont
  {X.}~\bibnamefont {Cai}}, \bibinfo {author} {\bibfnamefont {J.}~\bibnamefont
  {Peng}}, \bibinfo {author} {\bibfnamefont {S.}~\bibnamefont {Wang}}, \bibinfo
  {author} {\bibfnamefont {M.}~\bibnamefont {Tian}}, \bibinfo {author}
  {\bibfnamefont {Z.-A.}\ \bibnamefont {Xu}}, \bibinfo {author} {\bibfnamefont
  {W.}~\bibnamefont {Ku}}, \bibinfo {author} {\bibfnamefont {Z.}~\bibnamefont
  {Mao}}, \ and\ \bibinfo {author} {\bibfnamefont {Y.}~\bibnamefont {Liu}},\
  }\href {\doibase 10.1103/PhysRevB.97.041113} {\bibfield  {journal} {\bibinfo
  {journal} {Phys. Rev. B}\ }\textbf {\bibinfo {volume} {97}},\ \bibinfo
  {pages} {041113} (\bibinfo {year} {2018})}\BibitemShut {NoStop}%
\bibitem [{\citenamefont {Liu}(2011)}]{LiuGuo-Qiang2011}%
  \BibitemOpen
  \bibfield  {author} {\bibinfo {author} {\bibfnamefont {G.-Q.}\ \bibnamefont
  {Liu}},\ }\href {\doibase 10.1103/PhysRevB.84.235137} {\bibfield  {journal}
  {\bibinfo  {journal} {Phys. Rev. B}\ }\textbf {\bibinfo {volume} {84}},\
  \bibinfo {pages} {235137} (\bibinfo {year} {2011})}\BibitemShut {NoStop}%
\bibitem [{\citenamefont {Jin}\ and\ \citenamefont {Ku}(2018)}]{Zheting:2018}%
  \BibitemOpen
  \bibfield  {author} {\bibinfo {author} {\bibfnamefont {Z.}~\bibnamefont
  {Jin}}\ and\ \bibinfo {author} {\bibfnamefont {W.}~\bibnamefont {Ku}},\
  }\href@noop {} {\bibfield  {journal} {\bibinfo  {journal} {ArXiv e-prints}\ }
  (\bibinfo {year} {2018})},\ \Eprint {http://arxiv.org/abs/1809.04589}
  {arXiv:1809.04589 [cond-mat.mtrl-sci]} \BibitemShut {NoStop}%
\bibitem [{\citenamefont {Gou}\ \emph {et~al.}(2011)\citenamefont {Gou},
  \citenamefont {Grinberg}, \citenamefont {Rappe},\ and\ \citenamefont
  {Rondinelli}}]{Guo/Rondinelli:2011}%
  \BibitemOpen
  \bibfield  {author} {\bibinfo {author} {\bibfnamefont {G.}~\bibnamefont
  {Gou}}, \bibinfo {author} {\bibfnamefont {I.}~\bibnamefont {Grinberg}},
  \bibinfo {author} {\bibfnamefont {A.~M.}\ \bibnamefont {Rappe}}, \ and\
  \bibinfo {author} {\bibfnamefont {J.~M.}\ \bibnamefont {Rondinelli}},\
  }\href@noop {} {\bibfield  {journal} {\bibinfo  {journal} {Physical Review
  B}\ }\textbf {\bibinfo {volume} {84}},\ \bibinfo {pages} {144101} (\bibinfo
  {year} {2011})}\BibitemShut {NoStop}%
\bibitem [{\citenamefont {Grebinskij}\ \emph {et~al.}(2013)\citenamefont
  {Grebinskij}, \citenamefont {Masys}, \citenamefont
  {Mickevi\ifmmode~\check{c}\else \v{c}\fi{}ius}, \citenamefont {Lisauskas},\
  and\ \citenamefont {Jonauskas}}]{Grebinskij:2013}%
  \BibitemOpen
  \bibfield  {author} {\bibinfo {author} {\bibfnamefont {S.}~\bibnamefont
  {Grebinskij}}, \bibinfo {author} {\bibfnamefont {i.~c.~v.}\ \bibnamefont
  {Masys}}, \bibinfo {author} {\bibfnamefont {S.}~\bibnamefont
  {Mickevi\ifmmode~\check{c}\else \v{c}\fi{}ius}}, \bibinfo {author}
  {\bibfnamefont {V.}~\bibnamefont {Lisauskas}}, \ and\ \bibinfo {author}
  {\bibfnamefont {V.}~\bibnamefont {Jonauskas}},\ }\href {\doibase
  10.1103/PhysRevB.87.035106} {\bibfield  {journal} {\bibinfo  {journal} {Phys.
  Rev. B}\ }\textbf {\bibinfo {volume} {87}},\ \bibinfo {pages} {035106}
  (\bibinfo {year} {2013})}\BibitemShut {NoStop}%
\bibitem [{\citenamefont {Etz}\ \emph {et~al.}(2012)\citenamefont {Etz},
  \citenamefont {Maznichenko}, \citenamefont {B\"ottcher}, \citenamefont
  {Henk}, \citenamefont {Yaresko}, \citenamefont {Hergert}, \citenamefont
  {Mazin}, \citenamefont {Mertig},\ and\ \citenamefont {Ernst}}]{Etz:2012}%
  \BibitemOpen
  \bibfield  {author} {\bibinfo {author} {\bibfnamefont {C.}~\bibnamefont
  {Etz}}, \bibinfo {author} {\bibfnamefont {I.~V.}\ \bibnamefont
  {Maznichenko}}, \bibinfo {author} {\bibfnamefont {D.}~\bibnamefont
  {B\"ottcher}}, \bibinfo {author} {\bibfnamefont {J.}~\bibnamefont {Henk}},
  \bibinfo {author} {\bibfnamefont {A.~N.}\ \bibnamefont {Yaresko}}, \bibinfo
  {author} {\bibfnamefont {W.}~\bibnamefont {Hergert}}, \bibinfo {author}
  {\bibfnamefont {I.~I.}\ \bibnamefont {Mazin}}, \bibinfo {author}
  {\bibfnamefont {I.}~\bibnamefont {Mertig}}, \ and\ \bibinfo {author}
  {\bibfnamefont {A.}~\bibnamefont {Ernst}},\ }\href {\doibase
  10.1103/PhysRevB.86.064441} {\bibfield  {journal} {\bibinfo  {journal} {Phys.
  Rev. B}\ }\textbf {\bibinfo {volume} {86}},\ \bibinfo {pages} {064441}
  (\bibinfo {year} {2012})}\BibitemShut {NoStop}%
\bibitem [{\citenamefont {Akamatsu}\ \emph {et~al.}(2014)\citenamefont
  {Akamatsu}, \citenamefont {Fujita}, \citenamefont {Kuge}, \citenamefont
  {Sen~Gupta}, \citenamefont {Togo}, \citenamefont {Lei}, \citenamefont {Xue},
  \citenamefont {Stone}, \citenamefont {Rondinelli}, \citenamefont {Chen},
  \citenamefont {Tanaka}, \citenamefont {Gopalan},\ and\ \citenamefont
  {Tanaka}}]{Akamatsu:2014}%
  \BibitemOpen
  \bibfield  {author} {\bibinfo {author} {\bibfnamefont {H.}~\bibnamefont
  {Akamatsu}}, \bibinfo {author} {\bibfnamefont {K.}~\bibnamefont {Fujita}},
  \bibinfo {author} {\bibfnamefont {T.}~\bibnamefont {Kuge}}, \bibinfo {author}
  {\bibfnamefont {A.}~\bibnamefont {Sen~Gupta}}, \bibinfo {author}
  {\bibfnamefont {A.}~\bibnamefont {Togo}}, \bibinfo {author} {\bibfnamefont
  {S.}~\bibnamefont {Lei}}, \bibinfo {author} {\bibfnamefont {F.}~\bibnamefont
  {Xue}}, \bibinfo {author} {\bibfnamefont {G.}~\bibnamefont {Stone}}, \bibinfo
  {author} {\bibfnamefont {J.~M.}\ \bibnamefont {Rondinelli}}, \bibinfo
  {author} {\bibfnamefont {L.-Q.}\ \bibnamefont {Chen}}, \bibinfo {author}
  {\bibfnamefont {I.}~\bibnamefont {Tanaka}}, \bibinfo {author} {\bibfnamefont
  {V.}~\bibnamefont {Gopalan}}, \ and\ \bibinfo {author} {\bibfnamefont
  {K.}~\bibnamefont {Tanaka}},\ }\href {\doibase
  10.1103/PhysRevLett.112.187602} {\bibfield  {journal} {\bibinfo  {journal}
  {Phys. Rev. Lett.}\ }\textbf {\bibinfo {volume} {112}},\ \bibinfo {pages}
  {187602} (\bibinfo {year} {2014})}\BibitemShut {NoStop}%
\bibitem [{\citenamefont {Zhu}\ \emph {et~al.}(2017)\citenamefont {Zhu},
  \citenamefont {Cohen}, \citenamefont {Gibbs}, \citenamefont {Zhang},
  \citenamefont {Halasyamani}, \citenamefont {Hayward},\ and\ \citenamefont
  {Benedek}}]{zhu:2017}%
  \BibitemOpen
  \bibfield  {author} {\bibinfo {author} {\bibfnamefont {T.}~\bibnamefont
  {Zhu}}, \bibinfo {author} {\bibfnamefont {T.}~\bibnamefont {Cohen}}, \bibinfo
  {author} {\bibfnamefont {A.~S.}\ \bibnamefont {Gibbs}}, \bibinfo {author}
  {\bibfnamefont {W.}~\bibnamefont {Zhang}}, \bibinfo {author} {\bibfnamefont
  {P.~S.}\ \bibnamefont {Halasyamani}}, \bibinfo {author} {\bibfnamefont
  {M.~A.}\ \bibnamefont {Hayward}}, \ and\ \bibinfo {author} {\bibfnamefont
  {N.~A.}\ \bibnamefont {Benedek}},\ }\bibfield  {booktitle} {\emph {\bibinfo
  {booktitle} {Chemistry of Materials}},\ }\href {\doibase
  10.1021/acs.chemmater.7b03604} {\bibfield  {journal} {\bibinfo  {journal}
  {Chemistry of Materials}\ }\textbf {\bibinfo {volume} {29}},\ \bibinfo
  {pages} {9489} (\bibinfo {year} {2017})}\BibitemShut {NoStop}%
\bibitem [{\citenamefont {Fratino}\ \emph {et~al.}(2017)\citenamefont
  {Fratino}, \citenamefont {S\'emon}, \citenamefont {Charlebois}, \citenamefont
  {Sordi},\ and\ \citenamefont {Tremblay}}]{PhysRevB.95.235109}%
  \BibitemOpen
  \bibfield  {author} {\bibinfo {author} {\bibfnamefont {L.}~\bibnamefont
  {Fratino}}, \bibinfo {author} {\bibfnamefont {P.}~\bibnamefont {S\'emon}},
  \bibinfo {author} {\bibfnamefont {M.}~\bibnamefont {Charlebois}}, \bibinfo
  {author} {\bibfnamefont {G.}~\bibnamefont {Sordi}}, \ and\ \bibinfo {author}
  {\bibfnamefont {A.-M.~S.}\ \bibnamefont {Tremblay}},\ }\href {\doibase
  10.1103/PhysRevB.95.235109} {\bibfield  {journal} {\bibinfo  {journal} {Phys.
  Rev. B}\ }\textbf {\bibinfo {volume} {95}},\ \bibinfo {pages} {235109}
  (\bibinfo {year} {2017})}\BibitemShut {NoStop}%
\bibitem [{\citenamefont {Seki}\ \emph {et~al.}(2018)\citenamefont {Seki},
  \citenamefont {Shirakawa},\ and\ \citenamefont
  {Yunoki}}]{PhysRevB.98.205114}%
  \BibitemOpen
  \bibfield  {author} {\bibinfo {author} {\bibfnamefont {K.}~\bibnamefont
  {Seki}}, \bibinfo {author} {\bibfnamefont {T.}~\bibnamefont {Shirakawa}}, \
  and\ \bibinfo {author} {\bibfnamefont {S.}~\bibnamefont {Yunoki}},\ }\href
  {\doibase 10.1103/PhysRevB.98.205114} {\bibfield  {journal} {\bibinfo
  {journal} {Phys. Rev. B}\ }\textbf {\bibinfo {volume} {98}},\ \bibinfo
  {pages} {205114} (\bibinfo {year} {2018})}\BibitemShut {NoStop}%
\bibitem [{Note1()}]{Note1}%
  \BibitemOpen
  \bibinfo {note} {We use the Perdew-Burke-Ernzerhof exchange-correlation
  functional revised for solids (PBEsol) \cite {PBEsol:2008} as implemented in
  the Vienna \protect \textit {Ab initio} Simulation Package (VASP) \cite
  {Kresse/Furthmuller:1996b} with the projector augmented wave (PAW) approach
  \cite {Blochl/Jepsen/Andersen:1994} to treat the core and valence electrons
  using the following electronic configurations 3s$^2$3p$^6$4s$^2$ (Ca),
  4d$^7$5s$^1$ (Ru), 2s$^2$2p$^4$ (O) with a $5\times 5\times 3$ Monkhorst-Pack
  $k$-point mesh \cite {Monkhorst/Pack:1976} and a 650~eV planewave
  cutoff.}\BibitemShut {Stop}%
\bibitem [{\citenamefont {Zhang}\ and\ \citenamefont
  {Pavarini}(2017)}]{pavarini:2017}%
  \BibitemOpen
  \bibfield  {author} {\bibinfo {author} {\bibfnamefont {G.}~\bibnamefont
  {Zhang}}\ and\ \bibinfo {author} {\bibfnamefont {E.}~\bibnamefont
  {Pavarini}},\ }\href {\doibase 10.1103/PhysRevB.95.075145} {\bibfield
  {journal} {\bibinfo  {journal} {Phys. Rev. B}\ }\textbf {\bibinfo {volume}
  {95}},\ \bibinfo {pages} {075145} (\bibinfo {year} {2017})}\BibitemShut
  {NoStop}%
\bibitem [{\citenamefont {Sutter}\ \emph {et~al.}(2017)\citenamefont {Sutter},
  \citenamefont {Fatuzzo}, \citenamefont {Moser}, \citenamefont {Kim},
  \citenamefont {Fittipaldi}, \citenamefont {Vecchione}, \citenamefont
  {Granata}, \citenamefont {Sassa}, \citenamefont {Cossalter}, \citenamefont
  {Gatti}, \citenamefont {Grioni}, \citenamefont {R{\o}nnow}, \citenamefont
  {Plumb}, \citenamefont {Matt}, \citenamefont {Shi}, \citenamefont {Hoesch},
  \citenamefont {Kim}, \citenamefont {Chang}, \citenamefont {Jeng},
  \citenamefont {Jozwiak}, \citenamefont {Bostwick}, \citenamefont {Rotenberg},
  \citenamefont {Georges}, \citenamefont {Neupert},\ and\ \citenamefont
  {Chang}}]{SutterNatCom2017}%
  \BibitemOpen
  \bibfield  {author} {\bibinfo {author} {\bibfnamefont {D.}~\bibnamefont
  {Sutter}}, \bibinfo {author} {\bibfnamefont {C.~G.}\ \bibnamefont {Fatuzzo}},
  \bibinfo {author} {\bibfnamefont {S.}~\bibnamefont {Moser}}, \bibinfo
  {author} {\bibfnamefont {M.}~\bibnamefont {Kim}}, \bibinfo {author}
  {\bibfnamefont {R.}~\bibnamefont {Fittipaldi}}, \bibinfo {author}
  {\bibfnamefont {A.}~\bibnamefont {Vecchione}}, \bibinfo {author}
  {\bibfnamefont {V.}~\bibnamefont {Granata}}, \bibinfo {author} {\bibfnamefont
  {Y.}~\bibnamefont {Sassa}}, \bibinfo {author} {\bibfnamefont
  {F.}~\bibnamefont {Cossalter}}, \bibinfo {author} {\bibfnamefont
  {G.}~\bibnamefont {Gatti}}, \bibinfo {author} {\bibfnamefont
  {M.}~\bibnamefont {Grioni}}, \bibinfo {author} {\bibfnamefont {H.~M.}\
  \bibnamefont {R{\o}nnow}}, \bibinfo {author} {\bibfnamefont {N.~C.}\
  \bibnamefont {Plumb}}, \bibinfo {author} {\bibfnamefont {C.~E.}\ \bibnamefont
  {Matt}}, \bibinfo {author} {\bibfnamefont {M.}~\bibnamefont {Shi}}, \bibinfo
  {author} {\bibfnamefont {M.}~\bibnamefont {Hoesch}}, \bibinfo {author}
  {\bibfnamefont {T.~K.}\ \bibnamefont {Kim}}, \bibinfo {author} {\bibfnamefont
  {T.-R.}\ \bibnamefont {Chang}}, \bibinfo {author} {\bibfnamefont {H.-T.}\
  \bibnamefont {Jeng}}, \bibinfo {author} {\bibfnamefont {C.}~\bibnamefont
  {Jozwiak}}, \bibinfo {author} {\bibfnamefont {A.}~\bibnamefont {Bostwick}},
  \bibinfo {author} {\bibfnamefont {E.}~\bibnamefont {Rotenberg}}, \bibinfo
  {author} {\bibfnamefont {A.}~\bibnamefont {Georges}}, \bibinfo {author}
  {\bibfnamefont {T.}~\bibnamefont {Neupert}}, \ and\ \bibinfo {author}
  {\bibfnamefont {J.}~\bibnamefont {Chang}},\ }\href {\doibase
  10.1038/ncomms15176} {\bibfield  {journal} {\bibinfo  {journal} {Nature
  Communications}\ }\textbf {\bibinfo {volume} {8}},\ \bibinfo {pages} {15176}
  (\bibinfo {year} {2017})}\BibitemShut {NoStop}%
\bibitem [{\citenamefont {Das}\ \emph {et~al.}(2018)\citenamefont {Das},
  \citenamefont {Forte}, \citenamefont {Fittipaldi}, \citenamefont {Fatuzzo},
  \citenamefont {Granata}, \citenamefont {Ivashko}, \citenamefont {Horio},
  \citenamefont {Schindler}, \citenamefont {Dantz}, \citenamefont {Tseng},
  \citenamefont {McNally}, \citenamefont {R\o{}nnow}, \citenamefont {Wan},
  \citenamefont {Christensen}, \citenamefont {Pelliciari}, \citenamefont
  {Olalde-Velasco}, \citenamefont {Kikugawa}, \citenamefont {Neupert},
  \citenamefont {Vecchione}, \citenamefont {Schmitt}, \citenamefont {Cuoco},\
  and\ \citenamefont {Chang}}]{Das_Chang}%
  \BibitemOpen
  \bibfield  {author} {\bibinfo {author} {\bibfnamefont {L.}~\bibnamefont
  {Das}}, \bibinfo {author} {\bibfnamefont {F.}~\bibnamefont {Forte}}, \bibinfo
  {author} {\bibfnamefont {R.}~\bibnamefont {Fittipaldi}}, \bibinfo {author}
  {\bibfnamefont {C.~G.}\ \bibnamefont {Fatuzzo}}, \bibinfo {author}
  {\bibfnamefont {V.}~\bibnamefont {Granata}}, \bibinfo {author} {\bibfnamefont
  {O.}~\bibnamefont {Ivashko}}, \bibinfo {author} {\bibfnamefont
  {M.}~\bibnamefont {Horio}}, \bibinfo {author} {\bibfnamefont
  {F.}~\bibnamefont {Schindler}}, \bibinfo {author} {\bibfnamefont
  {M.}~\bibnamefont {Dantz}}, \bibinfo {author} {\bibfnamefont
  {Y.}~\bibnamefont {Tseng}}, \bibinfo {author} {\bibfnamefont {D.~E.}\
  \bibnamefont {McNally}}, \bibinfo {author} {\bibfnamefont {H.~M.}\
  \bibnamefont {R\o{}nnow}}, \bibinfo {author} {\bibfnamefont {W.}~\bibnamefont
  {Wan}}, \bibinfo {author} {\bibfnamefont {N.~B.}\ \bibnamefont
  {Christensen}}, \bibinfo {author} {\bibfnamefont {J.}~\bibnamefont
  {Pelliciari}}, \bibinfo {author} {\bibfnamefont {P.}~\bibnamefont
  {Olalde-Velasco}}, \bibinfo {author} {\bibfnamefont {N.}~\bibnamefont
  {Kikugawa}}, \bibinfo {author} {\bibfnamefont {T.}~\bibnamefont {Neupert}},
  \bibinfo {author} {\bibfnamefont {A.}~\bibnamefont {Vecchione}}, \bibinfo
  {author} {\bibfnamefont {T.}~\bibnamefont {Schmitt}}, \bibinfo {author}
  {\bibfnamefont {M.}~\bibnamefont {Cuoco}}, \ and\ \bibinfo {author}
  {\bibfnamefont {J.}~\bibnamefont {Chang}},\ }\href {\doibase
  10.1103/PhysRevX.8.011048} {\bibfield  {journal} {\bibinfo  {journal} {Phys.
  Rev. X}\ }\textbf {\bibinfo {volume} {8}},\ \bibinfo {pages} {011048}
  (\bibinfo {year} {2018})}\BibitemShut {NoStop}%
\bibitem [{\citenamefont {Dudarev}\ \emph {et~al.}(1998)\citenamefont
  {Dudarev}, \citenamefont {Botton}, \citenamefont {Savrasov}, \citenamefont
  {Humphreys},\ and\ \citenamefont {Sutton}}]{Dudarev/Sutton_et_al:1998}%
  \BibitemOpen
  \bibfield  {author} {\bibinfo {author} {\bibfnamefont {S.~L.}\ \bibnamefont
  {Dudarev}}, \bibinfo {author} {\bibfnamefont {G.~A.}\ \bibnamefont {Botton}},
  \bibinfo {author} {\bibfnamefont {S.~Y.}\ \bibnamefont {Savrasov}}, \bibinfo
  {author} {\bibfnamefont {C.~J.}\ \bibnamefont {Humphreys}}, \ and\ \bibinfo
  {author} {\bibfnamefont {A.~P.}\ \bibnamefont {Sutton}},\ }\href@noop {}
  {\bibfield  {journal} {\bibinfo  {journal} {Physical Review B}\ }\textbf
  {\bibinfo {volume} {57}},\ \bibinfo {pages} {1505} (\bibinfo {year}
  {1998})}\BibitemShut {NoStop}%
\bibitem [{sup()}]{suppmat}%
  \BibitemOpen
  \href@noop {} {}\bibinfo {note} {See Supplemental Material at [URL will be
  inserted by publisher] for additional structural data and electronic
  structures.}\BibitemShut {Stop}%
\bibitem [{\citenamefont {Puggioni}\ and\ \citenamefont
  {Rondinelli}(2014)}]{puggioni:2013}%
  \BibitemOpen
  \bibfield  {author} {\bibinfo {author} {\bibfnamefont {D.}~\bibnamefont
  {Puggioni}}\ and\ \bibinfo {author} {\bibfnamefont {J.~M.}\ \bibnamefont
  {Rondinelli}},\ }\href {\doibase 10.1038/ncomms4432} {\bibfield  {journal}
  {\bibinfo  {journal} {Nat. Commun.}\ }\textbf {\bibinfo {volume} {5}},\
  \bibinfo {pages} {3432} (\bibinfo {year} {2014})}\BibitemShut {NoStop}%
\bibitem [{\citenamefont {Lovesey}\ \emph {et~al.}(2019)\citenamefont
  {Lovesey}, \citenamefont {Khalyavin},\ and\ \citenamefont {van~der
  Laan}}]{Lovesey:2019}%
  \BibitemOpen
  \bibfield  {author} {\bibinfo {author} {\bibfnamefont {S.~W.}\ \bibnamefont
  {Lovesey}}, \bibinfo {author} {\bibfnamefont {D.~D.}\ \bibnamefont
  {Khalyavin}}, \ and\ \bibinfo {author} {\bibfnamefont {G.}~\bibnamefont
  {van~der Laan}},\ }\href {\doibase 10.1103/PhysRevB.99.134444} {\bibfield
  {journal} {\bibinfo  {journal} {Phys. Rev. B}\ }\textbf {\bibinfo {volume}
  {99}},\ \bibinfo {pages} {134444} (\bibinfo {year} {2019})}\BibitemShut
  {NoStop}%
\bibitem [{Note2()}]{Note2}%
  \BibitemOpen
  \bibinfo {note} {We use the Belov-Neronova-Smirnova (BNS) setting of magnetic
  space groups, see Bilbao Crystallographic server, \protect \href
  {http://www.cryst.ehu.es}{http://www.cryst.ehu.es}.}\BibitemShut {Stop}%
\bibitem [{\citenamefont {Liu}\ \emph {et~al.}(1999)\citenamefont {Liu},
  \citenamefont {Yoon}, \citenamefont {Cooper}, \citenamefont {Cao},\ and\
  \citenamefont {Crow}}]{PhysRevB.60.R6980}%
  \BibitemOpen
  \bibfield  {author} {\bibinfo {author} {\bibfnamefont {H.~L.}\ \bibnamefont
  {Liu}}, \bibinfo {author} {\bibfnamefont {S.}~\bibnamefont {Yoon}}, \bibinfo
  {author} {\bibfnamefont {S.~L.}\ \bibnamefont {Cooper}}, \bibinfo {author}
  {\bibfnamefont {G.}~\bibnamefont {Cao}}, \ and\ \bibinfo {author}
  {\bibfnamefont {J.~E.}\ \bibnamefont {Crow}},\ }\href {\doibase
  10.1103/PhysRevB.60.R6980} {\bibfield  {journal} {\bibinfo  {journal} {Phys.
  Rev. B}\ }\textbf {\bibinfo {volume} {60}},\ \bibinfo {pages} {R6980}
  (\bibinfo {year} {1999})}\BibitemShut {NoStop}%
\bibitem [{\citenamefont {Zhang}\ and\ \citenamefont
  {Pavarini}(2018)}]{Zhang2018}%
  \BibitemOpen
  \bibfield  {author} {\bibinfo {author} {\bibfnamefont {G.}~\bibnamefont
  {Zhang}}\ and\ \bibinfo {author} {\bibfnamefont {E.}~\bibnamefont
  {Pavarini}},\ }\href {\doibase 10.1002/pssr.201800211} {\bibfield  {journal}
  {\bibinfo  {journal} {physica status solidi ({RRL}) - Rapid Research
  Letters}\ }\textbf {\bibinfo {volume} {12}},\ \bibinfo {pages} {1800211}
  (\bibinfo {year} {2018})}\BibitemShut {NoStop}%
\bibitem [{\citenamefont {McIntosh}\ and\ \citenamefont
  {Kaiser}(1996)}]{McIntosh:1996}%
  \BibitemOpen
  \bibfield  {author} {\bibinfo {author} {\bibfnamefont {G.~C.}\ \bibnamefont
  {McIntosh}}\ and\ \bibinfo {author} {\bibfnamefont {A.~B.}\ \bibnamefont
  {Kaiser}},\ }\href {\doibase 10.1103/PhysRevB.54.12569} {\bibfield  {journal}
  {\bibinfo  {journal} {Phys. Rev. B}\ }\textbf {\bibinfo {volume} {54}},\
  \bibinfo {pages} {12569} (\bibinfo {year} {1996})}\BibitemShut {NoStop}%
\bibitem [{\citenamefont {Perdew}\ \emph {et~al.}(2008)\citenamefont {Perdew},
  \citenamefont {Ruzsinszky}, \citenamefont {Csonka}, \citenamefont {Vydrov},
  \citenamefont {Scuseria}, \citenamefont {Constantin}, \citenamefont {Zhou},\
  and\ \citenamefont {Burke}}]{PBEsol:2008}%
  \BibitemOpen
  \bibfield  {author} {\bibinfo {author} {\bibfnamefont {J.~P.}\ \bibnamefont
  {Perdew}}, \bibinfo {author} {\bibfnamefont {A.}~\bibnamefont {Ruzsinszky}},
  \bibinfo {author} {\bibfnamefont {G.~I.}\ \bibnamefont {Csonka}}, \bibinfo
  {author} {\bibfnamefont {O.~A.}\ \bibnamefont {Vydrov}}, \bibinfo {author}
  {\bibfnamefont {G.~E.}\ \bibnamefont {Scuseria}}, \bibinfo {author}
  {\bibfnamefont {L.~A.}\ \bibnamefont {Constantin}}, \bibinfo {author}
  {\bibfnamefont {X.}~\bibnamefont {Zhou}}, \ and\ \bibinfo {author}
  {\bibfnamefont {K.}~\bibnamefont {Burke}},\ }\href@noop {} {\bibfield
  {journal} {\bibinfo  {journal} {Physical Review Letters}\ }\textbf {\bibinfo
  {volume} {100}},\ \bibinfo {pages} {136406} (\bibinfo {year}
  {2008})}\BibitemShut {NoStop}%
\bibitem [{\citenamefont {Kresse}\ and\ \citenamefont
  {Furthm\"uller}(1996)}]{Kresse/Furthmuller:1996b}%
  \BibitemOpen
  \bibfield  {author} {\bibinfo {author} {\bibfnamefont {G.}~\bibnamefont
  {Kresse}}\ and\ \bibinfo {author} {\bibfnamefont {J.}~\bibnamefont
  {Furthm\"uller}},\ }\href@noop {} {\bibfield  {journal} {\bibinfo  {journal}
  {Computational Materials Science}\ }\textbf {\bibinfo {volume} {6}},\
  \bibinfo {pages} {15 } (\bibinfo {year} {1996})}\BibitemShut {NoStop}%
\bibitem [{\citenamefont {Bl\"ochl}\ \emph {et~al.}(1994)\citenamefont
  {Bl\"ochl}, \citenamefont {Jepsen},\ and\ \citenamefont
  {Andersen}}]{Blochl/Jepsen/Andersen:1994}%
  \BibitemOpen
  \bibfield  {author} {\bibinfo {author} {\bibfnamefont {P.~E.}\ \bibnamefont
  {Bl\"ochl}}, \bibinfo {author} {\bibfnamefont {O.}~\bibnamefont {Jepsen}}, \
  and\ \bibinfo {author} {\bibfnamefont {O.~K.}\ \bibnamefont {Andersen}},\
  }\href@noop {} {\bibfield  {journal} {\bibinfo  {journal} {Physical Review
  B}\ }\textbf {\bibinfo {volume} {49}},\ \bibinfo {pages} {16223} (\bibinfo
  {year} {1994})}\BibitemShut {NoStop}%
\bibitem [{\citenamefont {Monkhorst}\ and\ \citenamefont
  {Pack}(1976)}]{Monkhorst/Pack:1976}%
  \BibitemOpen
  \bibfield  {author} {\bibinfo {author} {\bibfnamefont {H.~J.}\ \bibnamefont
  {Monkhorst}}\ and\ \bibinfo {author} {\bibfnamefont {J.~D.}\ \bibnamefont
  {Pack}},\ }\href@noop {} {\bibfield  {journal} {\bibinfo  {journal} {Physical
  Review B}\ }\textbf {\bibinfo {volume} {13}},\ \bibinfo {pages} {5188}
  (\bibinfo {year} {1976})}\BibitemShut {NoStop}%
\end{thebibliography}%

\end{document}